\documentclass[
    ,final            % use final for the camera ready runs
  ]
  {aipproc}

\layoutstyle{6x9}

\begin{document}

\title{Low Energy Antiproton Experiments -- A Review}


\author{Klaus P. Jungmann}{
  address={Kernfysisch Versneller Instituut, Rijksuniversiteit Groningen,  Zernikelaan 25\\ 
           9747AA Groningen, The Netherlands}
}

%\author{<author2>}{
%  address={<common address for author2 and author3>}
%}
%
%\author{<author3>}{
%  address={<common address for author2 and author3>}
%  ,altaddress={<author1 address>} % additional visiting address
%}

\begin{abstract}
Low energy antiprotons offer excellent opportunities to study properties of 
fundamental forces and symmetries in nature. Experiments with them can contribute 
substantially to deepen our fundamental knowledge in atomic, nuclear and 
particle physics. Searches for new interactions can be carried out 
by studying discrete symmetries. Known interactions can be tested 
precisely and fundamental constants can be extracted from accurate 
measurements on free antiprotons ($\overline{p}$'s) and bound two- and three-body systems such
as antihydrogen ($\overline{{\rm H}}=\overline{p}e^-$), the antprotonic helium ion
(He$^{++} \overline{p}$)$^+$ and the antiprotonic atomcule (He$^{++} \overline{p}e^-$) . 
The trapping of a single  $\overline{p}$ in a Penning trap, the formation and precise studies 
of antiprotonic helium ions and atoms and recently the production of 
$\overline{{\rm H}}$  have been among the pioneering experiments. 
They have led already to precise values for $\overline{p}$  
parameters, accurate tests of bound two- and three-body Quantum Electrodynamics (QED), 
tests of the CPT theorem  and a  better understanding of atom formation 
from their constituents. 
Future experiments promise more precise tests of the standard theory and 
have a robust potential to discover new physics. 
Precision experiments with low energy $\overline{p}$'s share the need for intense 
particle sources and the need for time to develop novel instrumentation with 
all other experiments, which aim for high precision in exotic fundamental 
systems. The experimental programs - carried out in the past mostly at the 
former LEAR  facility and at present at the AD facility at CERN - 
would benefit from intense future sources of low energy $\overline{p}$'s. The 
highest possible $\overline{p}$ fluxes should be aimed for at new facilities
such as the planned FLAIR facility at GSI in order to maximize the potential
of delicate precision experiments to influence model building.  
Examples of key $\overline{p}$ experiments are discussed here and compared 
with other experiments in the field. Among the central issues is 
their potential  to obtain important information on basic symmetries such 
as CPT and to gain insights into antiparticle gravitation as well as the 
possibilities to learn about nuclear neutron distributions.
\end{abstract}

\maketitle

%%%%%%%%%%%%%%%%%%%%%%%%%%%%%%%%%%%%%%%%%%%%
%% MAINMATTER
%%%%%%%%%%%%%%%%%%%%%%%%%%%%%%%%%%%%%%%%%%%%

\section{Introduction}

The availability of beams of low energy antiprotons ($\overline{p}$) has led 
to a number of precision experiments on the properties of free $\overline{p}$'s,
of simple two- and three-body bound systems and to studies concerning the
interactions of $\overline{p}$'s with matter.
Central to the motivation for the precision experiments is the goal to test
fundamental forces and symmetries in physics, 
in particular the CPT invariance \cite{CPT_0, PDG_2004}.  

 To date we know four fundamental interactions:
(i)    Electromagnetism, 
(ii)     Weak Interactions,
(iii)    Strong Interactions, and 
(iv)   Gravitation.
These forces are considered fundamental, because all
observed dynamical processes in physics can be traced back to one or
a combination of them. Together with fundamental symmetries they from a framework on
which all physical descriptions ultimately rest.
The Electromagnetic, the Weak and many aspects of the Strong 
Interactions can be described to astounding precision in one single 
coherent picture, the Standard Model (SM). Gravity is not 
included  in the SM.
It is a major goal in modern physics to find a unified 
quantum field theory which includes all the four known
fundamental forces in physics. A satisfactory quantum 
description of gravity remains yet to 
be found.
% and is a lively field of actual activity. 

In modern physics - and in particular in the SM -
symmetries play an important and central role. Where\-as
global symmetries relate to conservation laws, local symmetries 
yield forces \cite{Lee_56}. 
It is rather unsatisfactory that within the SM
the physical origin of the observed breaking of discrete 
symmetries in weak interactions,
e.g. of parity (P), of time reversal (T) and of 
combined charge conjugation and parity (CP), 
remains unrevealed, although the experimental findings can be well
described. Further, there are observed conservation laws which
have no status in physics, as there is no symmetry known from which
they could be derived, e.g. conservation of baryon and lepton numbers.

Among the intriguing questions in modern physics are the hierarchy of the 
fundamental fermion masses and 
the number of fundamental particle generations.
Further, the electro-weak SM has a rather large number of some 
27 free parameters. All of them need to be extracted from experiments.

The spectrum of speculative models beyond the %present 
standard theory,
which try to expand the SM,
includes such which involve left-right symmetry, 
fundamental fermion compositeness, new particles, leptoquarks, 
supersymmetry, supergravity and many more. Interesting candidates 
for an all encompassing quantum field theory are string or membrane
(M) theories which in their low energy limit include supersymmetry.

Searches for effects predicted in speculative models can be carried out
in accelerator experiments at the highest at presently possible energies, where, e.g., 
the direct production of new particles can be  looked for. Complementary,
at low energies one can look for small deviations from accurate
predictions within the SM in high precision experiments. 
Low energy antiproton experiments are important examples of
such precision experiments. In this article we describe precision 
low energy antiproton experiments and compare them in physics potential and 
experimental techniques with related  low energy precision experiments in other
systems.

\section{Antiproton Sources}
%\subsection{Present Situation}

Today there exists only one facility for low energy antiproton research,
the Antiproton Decelerator (AD) at CERN. It succeeded the former LEAR facility
and  delivers about $10^5 $ $\overline{p}$/s and 
has only pulsed extraction ($3\times 10^7$  $\overline{p}$ 
in a 100 ns long pulse every 85 seconds).  The AD output energy  is 5 MeV. 
This relatively high kinetic energy is a major disadvantage of the AD . For the 
experimental programme energies below 100 keV would be much better suited.
To achieve this, typically degrader foils are used and one obtains about 2500
$\overline{p}$/s per AD pulse.
The AD  serves three experimental areas occupied by the ASACUSA, the ATHENA
and the ATRAP collaborations. The AD serves  a physics program 
including the spectroscopy of antiprotonic atoms and $\overline{{\rm H}}$ and investigations 
of the $\overline{p}$ interaction with matter.

\begin{figure}[t]
  \includegraphics[height=.33\textheight,clip]{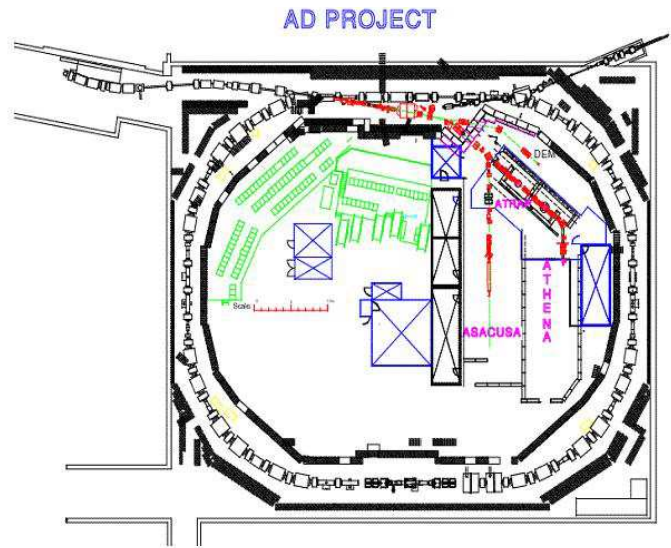}
  \caption{At present only the Antiproton Decelerator (AD) facility
           at CERN provides low energy $\overline{p}$'s for precision experiments.
           A possible upgrade with the ELENA storage ring could increase the
            $\overline{{\rm p}}$ flux significantly. In the long term future
           the FLAIR facility at GSI could provide low energy 
           $\overline{{\rm p}}$'s for a variety of experiments at somewhat 
           higher rates    
               }
\end{figure}
 
Recently an ultra slow monoenergetic $\overline{p}$ beam could be extracted 
with relatively high efficiency.  A radiofrequency quadrupole (RFQ) decelerator was employed 
to achieve  $1.2 \times 10^6$ particles per AD pulse could be obtained. When loaded
into a  multiring trap (MRT) and with electron cooling beams of 10-500eV
energy could be made \cite{Yamazaki_2005}.  
This new development presents a major step forward and could open  new research fields 
with ultra slow $\overline{p}$'s, in particular precision tests of fundamental interactions.
\begin{figure}[b]
  \includegraphics[height=.19\textheight,clip]{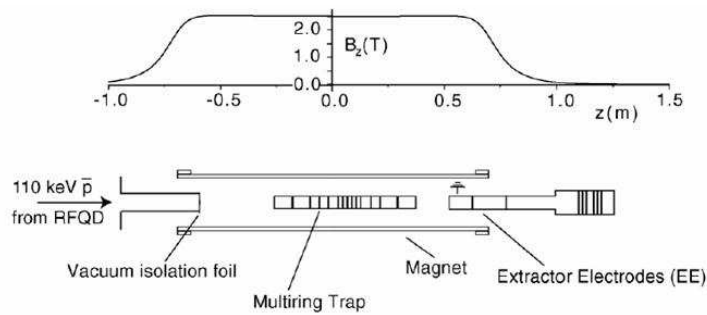}
  \caption{The multiring trap at the CERN AD facility which achieved  10-500 eV 
                  $\overline{p}$ beams \cite{Yamazaki_2005}.
               }
\end{figure}

%\subsection{Future Possibilities}
 At CERN a new Extra Low ENergy
Antiproton ring (ELENA) has been proposed, which would be installed behind  the AD. One
expects 300 ns wide bunches in the energy range 5.3 MeV to 0.1 MeV. Slow extraction 
appears possible. The space 
charge limit of the device is about $1.7 \times 10^7$  particles; the 
longitudinal and transverse 
temperatures are expected in the 100 eV range.

\begin{figure}[t]
  \includegraphics[height=.27\textheight,clip]{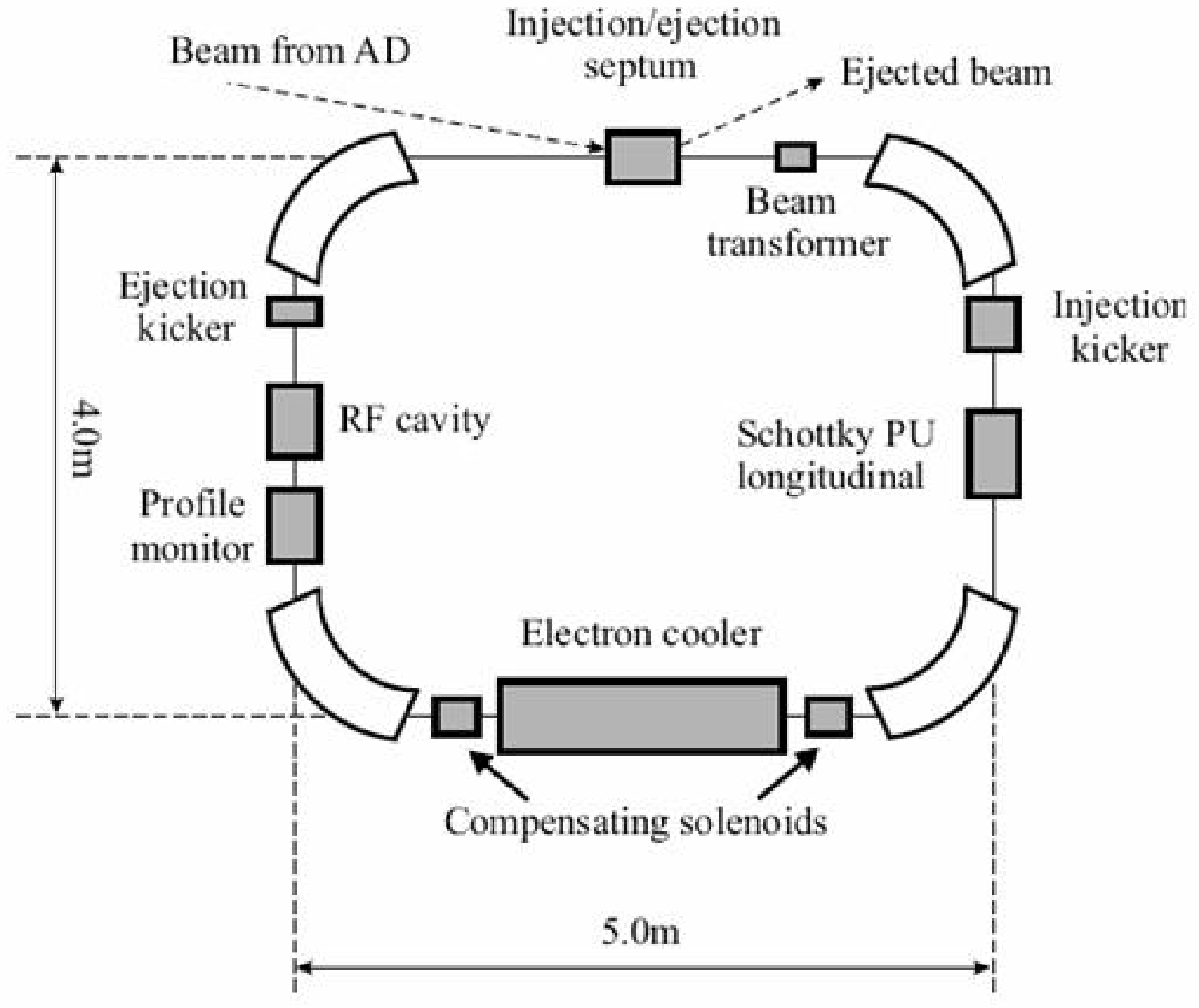}  
  \includegraphics[height=.27\textheight,clip]{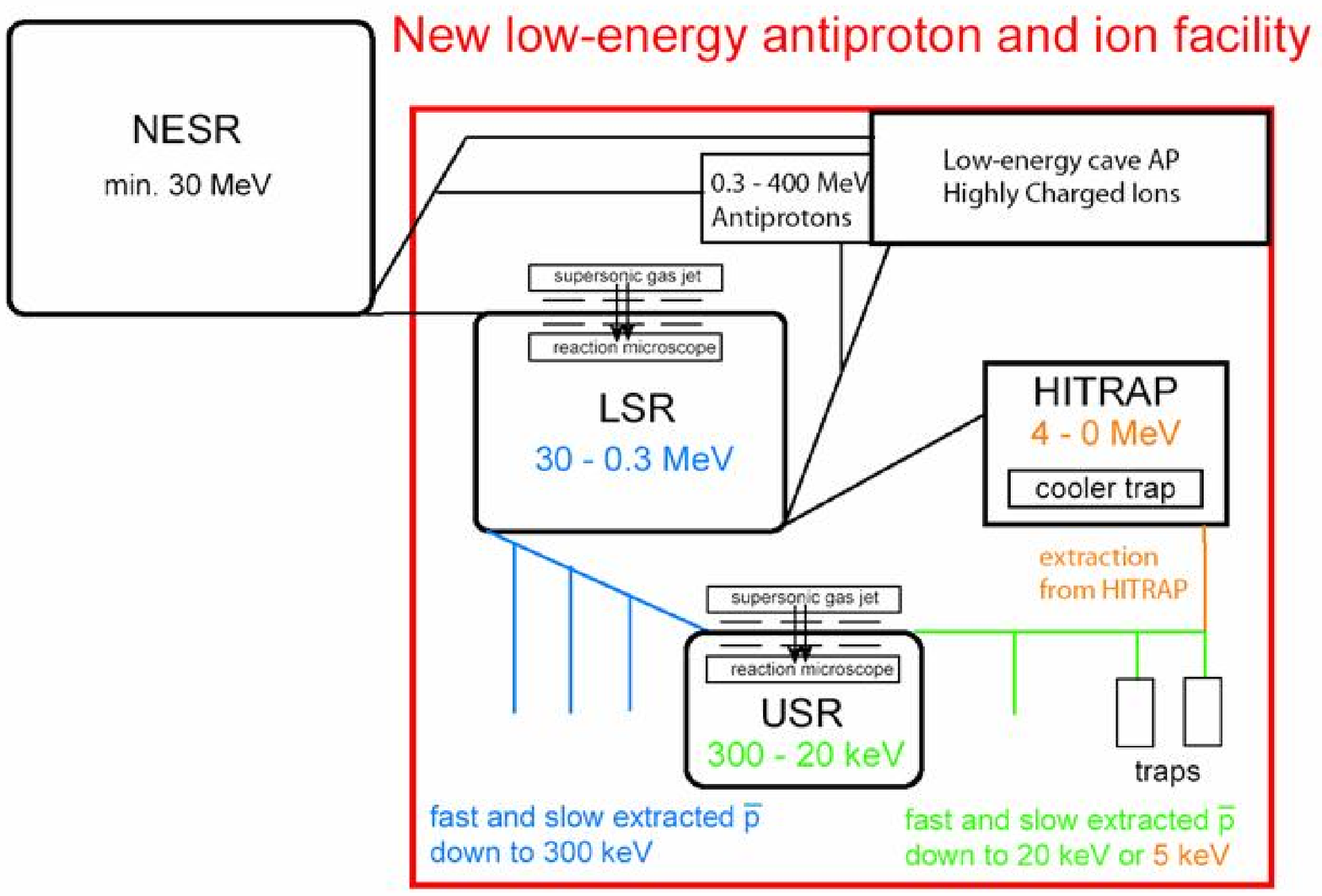} 
\caption{ A possible upgrade of the CERN AD facility with the 
           LENA  ring (left) \cite{ELENA_2004} could increase the
            $\overline{{\rm p}}$ flux significantly. In the long term future
           the FLAIR facility (right) at FAIR(GSI)  \cite{FLAIR_2005}
           could provide low energy 
           $\overline{{\rm p}}$'s for a variety of experiments at somewhat 
           higher rates.    
               }
\end{figure} 

In connection with the future upgrade plans of GSI, Germany, into a 
Facility Antiproton and Ion Research (FAIR)  also a 
Facility for Low-energy Antiproton and Ion Research (FLAIR) is foreseen.
This promises cooled $\overline{p}$ beams of $1 \times 10^6$ $\overline{p}$/s at 300 keV,
and $5 \times  10^5$ $\overline{p}$/s at 20 keV. Emittances of 1$\pi$ mm mrad and a 
momentum spread of  $< 10^{-3}$  are possible which is a significant improvement
in brilliance over the present AD facility.

%\section{Fundamental systems in Atomic Physics}

\section{Properties of Elementary Particles -- Stored and Trapped Leptons and Baryons}

\begin{figure}[b]
  \includegraphics[height=.23\textheight,clip]{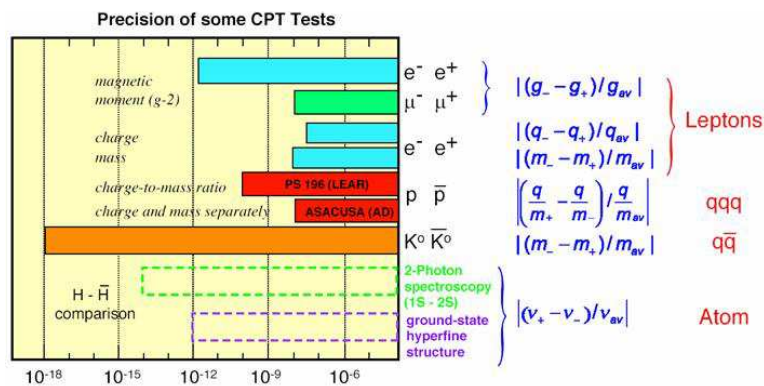}
  \caption{\label{CPT_tests} 
                  Tests of the CPT theorem comparing particle and antiparticle
                  properties. Reported is the limit of the fractional differences achieved. 
                  Also shown are
                  expectations from $\overline{{\rm H}}$ experiments \cite{Widmann_2005}.}
\end{figure}

The properties of elementary particles offer opportunities to test standard theory, 
and to search for new physics.  Particularly the comparisons of
particle and antiparticle properties have often been reported as tests of the
CPT theorem, which predicts that except for the sign of their electric charge 
(and charge related quantities) particles and antiparticles should be identical
(see Fig. \ref{CPT_tests}).

Trapping and storing of charged particles in combined magnetic and electric fields
has been very successfully applied for obtaining properties of 
the respective species and for determining most accurate values of fundamental constants.
Most accurate results were obtained from single trapped and cooled charged particles.
The comparison of proton ($p$)  and $\overline{p}$ has already reached an impressive
level of precision (see Fig. \ref{CPT_tests}).

\subsection{Leptons}

\begin{figure}[b]
  \includegraphics[height=.27\textheight,clip]{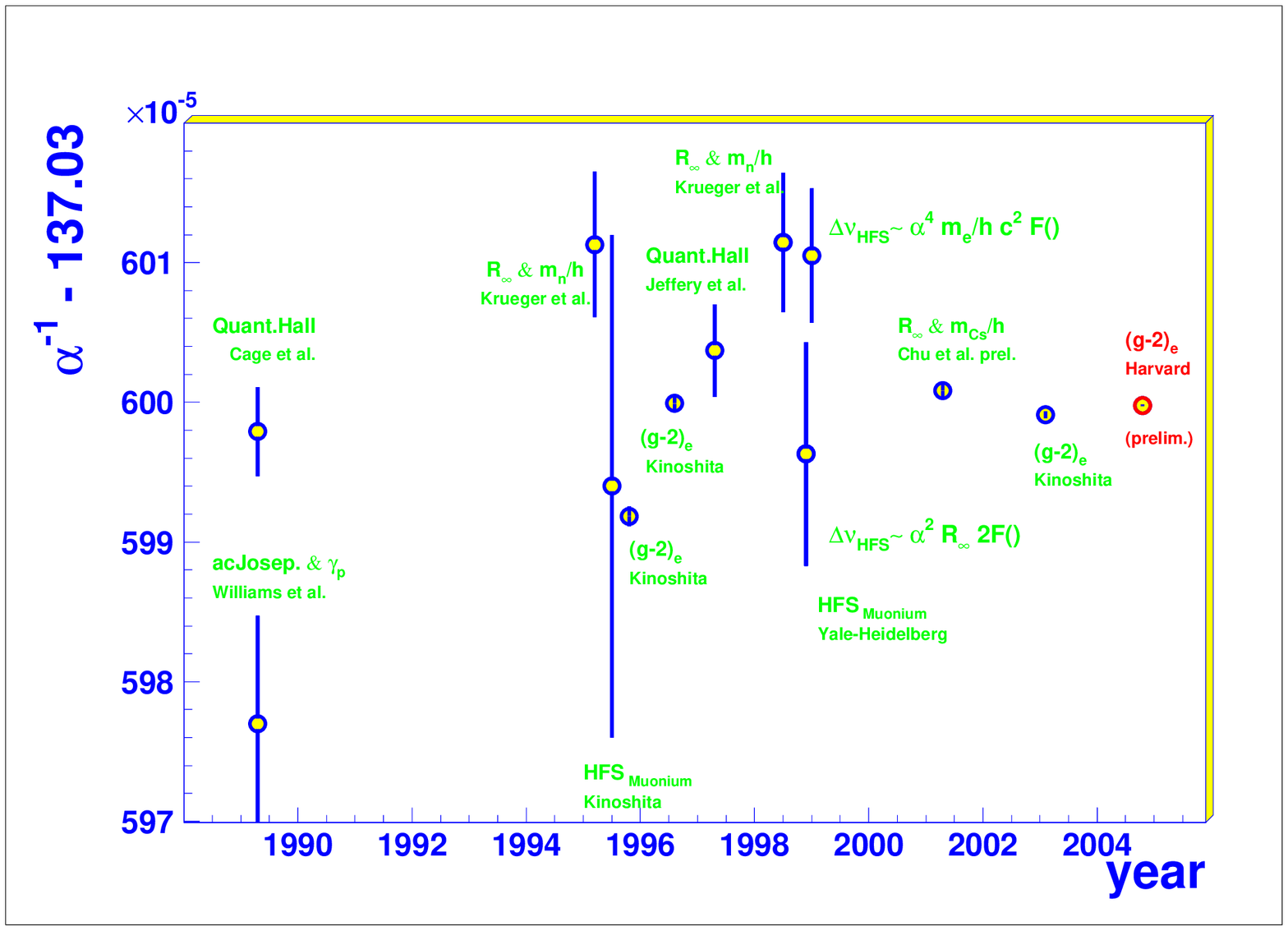}
  \caption{%Various determinations of the fine structure constant $\alpha$.
                  \label{ALPHA}
                  Measurements of the electron magnetic anomaly 
                  using single electrons confined and cooled  in a Penning trap yield the best
                  values for the fine structure constant $\alpha$.}
\end{figure}

The magnetic anomaly of fermions $a=\frac{1}{2}\cdot(g-2)$
describes the deviation of their magnetic g-factor
from the value 2 predicted in the Dirac theory. It
could be determined for single electrons and positrons in Penning traps
by Dehmelt and his coworkers to 10~ppb \cite{Dyc_90}
by measuring the cyclotron frequency and its difference to the 
spin precession frequency (g-2 measurement).
The good agreement for the magnetic anomaly for electrons and positrons is
considered the best CPT test for leptons \cite{PDG_2004}.  
Accurate calculations involving
almost exclusively the "pure" QED
of electron, positron and photon fields allow the most precise
determination of the fine structure constant $\alpha$ \cite{Kinoshita_2004,Kinoshita_1990}
by comparing experiment and theory for the electron magnetic anomaly
in which $\alpha$ appears as an expansion coefficient (see Fig. \ref{ALPHA}).
One order of magnitude improvement appears  possible \cite{Gabrielse_2005}
with a new experimental approach (also involving single cooled trapped particles) 
which aims for
reducing the effect of cavity QED, the major systematic contribution to the 
previous experiment \cite{Brown_1986}.

Muons have been stored in series of measurements at CERN and BNL in magnetic 
storage rings with weak electrostatic focusing, which is conceptually
equivalent to Penning traps.
The latest experimental results \cite{Bennet_2004}
yield  values for positive and negative muons which agree
at the 0.7 ppm level in agreement with CPT. 
The muon is by a factor $(\frac{m_{\mu}}{m_e})^2$  more sensitive
to heavy particles compared to the electron. The muon g-2 measurements are
sensitive to new physics involving heavy particles
at 40,000 times lower experimental precision. Whether
the present experimental results are in agreement with standard theory remains an
open question, as not sufficiently accurate values for corrections due to known
strong interaction effects exist yet.

\subsection{Protons and Antiprotons}

After moderation of a $\overline{p}$ beam from LEAR at CERN, $\overline{p}$'s
could be trapped in a cylindrical Penning trap for the first time in 1986 \cite{Gabrielse_1986}.
Effective moderation of MeV  $\overline{p}$'s  and their capture were very important, 
which has led to detailed  studies of the range differences when  $p$'s and $\overline{p}$'s 
are slowed down in matter,  known as the Barkas effect\cite{Gabrielse_1989_a}.  
Further electron cooling is essential and could be demonstrated already in the early 
experiments \cite{Kells_1986}.

\begin{figure}[h]
  \includegraphics[height=.25\textheight,clip]{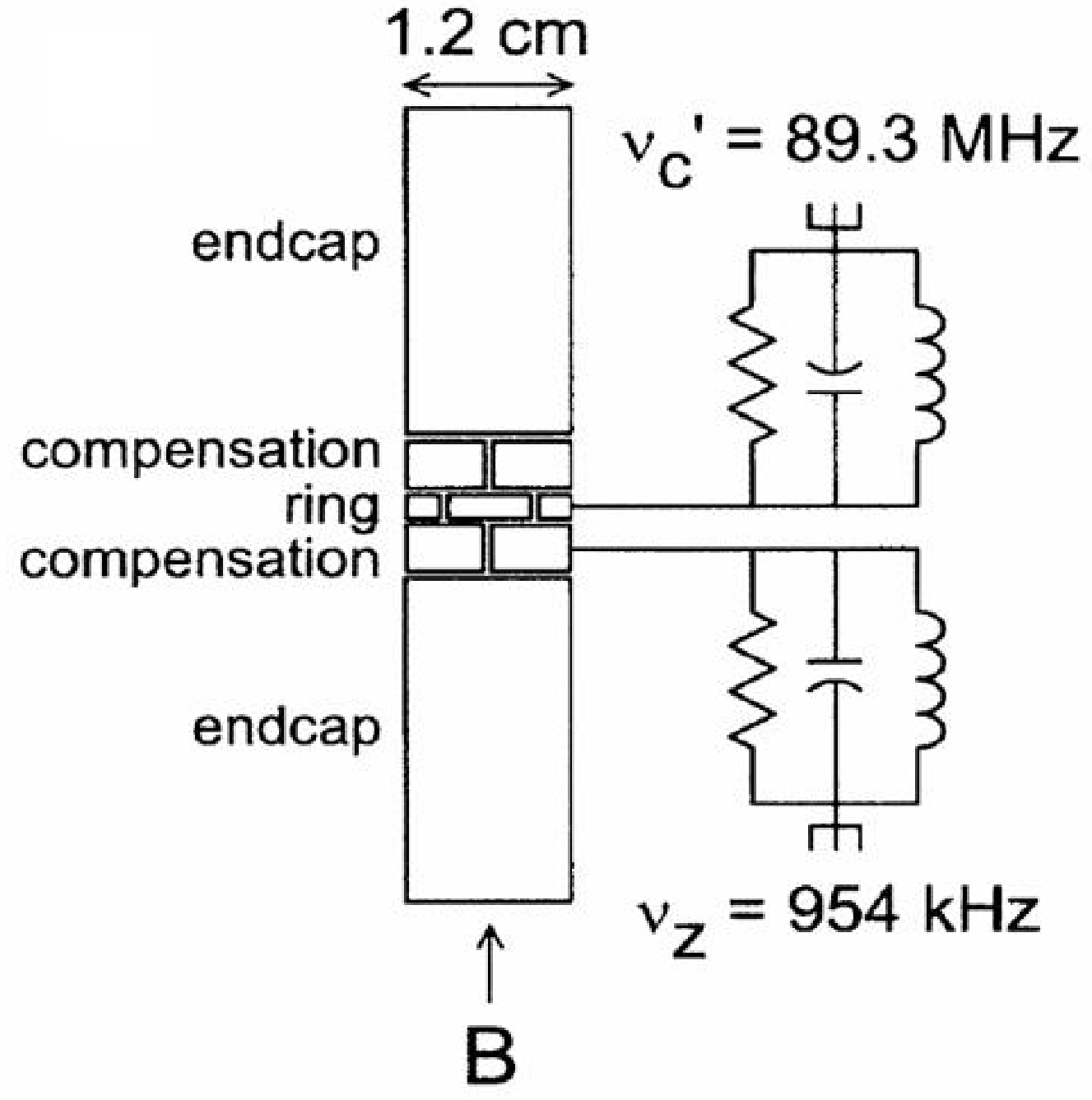}
  \includegraphics[height=.25\textheight,clip]{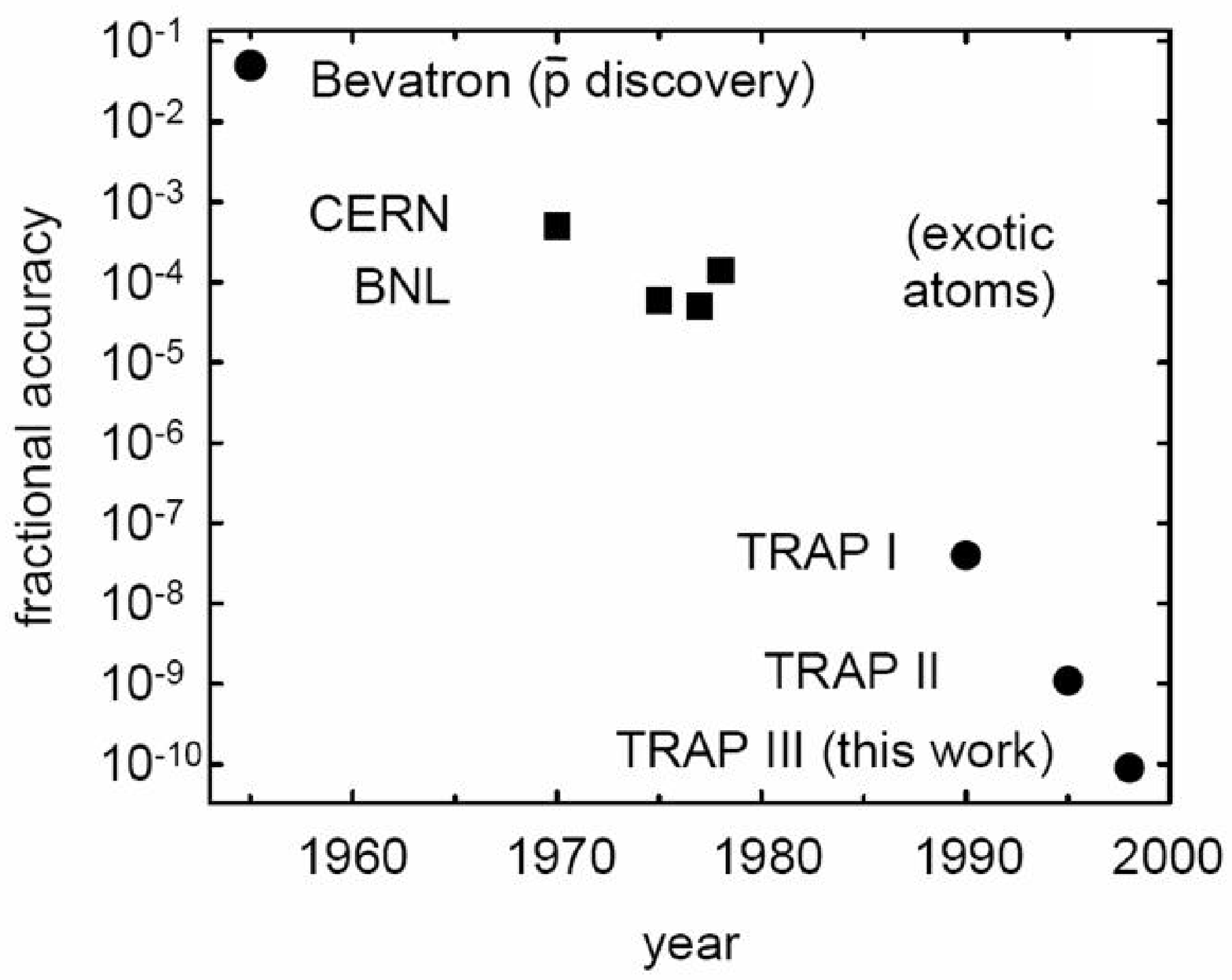}
  \caption{\label{e_m_pbar}
                   Left: Cylindrical Penning trap for $\overline{p}$'s \cite{Gabrielse_1986}. Right:
                   The charge to mass ratio of protons and $\overline{p}$'s 
                   could be measured from the 
                   cyclotron frequency in a Pennig trap.
                   The fractional uncertainty was  constantly improved.
                   The best value was reached, when a single $\overline{p}$ was stored and compared
                   to H$^-$ ions \cite{Gabrielse_1995,Gabrielse_1999}.}
\end{figure}

In a series of measurements in
which the cyclotron frequencies were measured the accuracy of the charge to mass ratio
for $\overline{p}$'s could be improved (see Fig. \ref{e_m_pbar})  and compared to the proton
value. The best results were achieved when a single H$^-$ ion and 
a single $\overline{p}$ where measured alternatively in the same trap \cite{Gabrielse_1999}.
%It should be mentioned that even for cooled particles relativistic corrections play a role at the
%achieved level of accuracy. 
At present these experiments are interpreted as a CPT test for
$p$ and $\overline{p}$ at the level of $9\times 10^{-11}$.
A new experiment has been proposed to measure  the magnetic g-factor of the
$\overline{p}$ using single particle trapping. Similar to measurements on
single electrons and positrons the cyclotron and spin precession frequencies 
shall be determined. One expects for the comparison of $p$ and $\overline{p}$
g-factors an improvement by a factor of $10^6$ \cite{FLAIR_2005}.

\section{Hydrogen-like Atoms}

Next to single particles, H-like atoms ( see Table \ref{tab_H}) are the simplest systems
in atomic physics. The electromagnetic part of their binding can be 
calculated to very high precision in the framework of QED
\cite{Kinoshita_1990,Eides_2001}. This allows to study the influence of other known
interactions in the Standard Model such as strong and weak forces. Also new yet
unknown interactions beyond the Standard Model can be searched for when the precise 

\begin{table}[!t]
 \caption{\label{tab_H}
    The ground state hyperfine structure splitting $\Delta \nu_{HFS}$
    and the 1S-2S level separation $\Delta \nu_{1S-2S}$
    of H and  some
    exotic H-like systems offer narrow transitions for
    studying the interactions in Coulomb bound two-body systems.
    In the exotic systems the natural linewidths $\delta \nu_{HFS}$ and $\delta \nu_{1S-2S}$ 
    have a fundamental lower limit given by the finite lifetime of
    the systems, because of annihilation, as in the case of positronium and antiprotonic helium,
    or because of weak muon or pion decays.
    The very high
    quality factors (transition frequency divided by the natural linewidth
    $\Delta \nu / \delta \nu$ ) in
    H and other systems with hadronic nuclei can unfortunately not
    be fully utilized to test
    the theory, because of the insufficiently known charge distribution and
    dynamics of the charge carrying constituents within the hadrons. The purely leptonic
    systems are not affected by this complication. 
    }
 \vspace{2mm}
 {
 \begin{tabular}[!t]{|c|*{9}{r@{}l|}}
\hline
 &&{}    &&{}    &&{}    &&{}    &&{}  &&{}    &&{} &&{} &&{} \\
 &Posit&{ronium}  &Muon&{ium}   &Hydr&{ogen~~/}  &Muo&{nic}
 &Pion&{ium}	    &Muo&{nic}  &Pion&{ic} & Antipro&{tonic} &Prot&{onium} \\
 &&  &&   &Antihy&{drogen}  &Heliu&{m4}  
 &&	    &Hydr&{ogen} &Hydr&{ogen} &Helium&{4 ion} &&{} \\
 &&{}    &&{}    &&{}    &&{}    &&{}  &&{}  &&{}   &&{} &&{} \\
 &{ $e^+$}&{{$e^-$}}
 &{$\mu^+$}&{{$e^-$}}
 &{$p^{ }$}{$e^-$}&{~~/~~~{$\overline{p}$}{$e^+$}}
 &{ $(\alpha\mu^-$}&{{$)e^-$}}
 &{$\pi^+$}&{{$e^-$}}
 &{$ p^{ }$}&{{$\mu^-$}} 
 &{$ p^{ }$}&{{$\pi^-$}} 
 &{$\alpha$}&{{$\overline{p}$}}
 &{$p^{ }$}&{{$\overline{p}$}} \\
 &&{}    &&{}    &&{}    &&{}    &&{}    &&{} &&{}  &&{} &&{} \\
\hline
%\hline
%
  &&{}    &&{}    &&{}   &&{}  &&{}    &&{}  &&{}    &&{} &&{} \\
{{$\Delta\nu _{1S-2S}$}} &123&{3.6}$$
				  &245&{5.6}   &246&{6.1}  &246&{8.5}
 &245&{8.6}	    &4.59&{$\times 10^5$}  &5.88&{$10^5$} &9&{$\times 10^7$} 
 &2.25&{$\times 10^7$} \\
$[$THz$]$	&&{}    &&{}    &&{}    &&{}    &&{}    &&{}   &&{} &&{}  &&{} \\
%%\hline
	&&{}    &&{}    &&{}    &&{}    &&{}    &&{}  &&{} &&{} &&{} \\
{{$\delta \nu_{1S-2S}$}} &1&{.28}$^{\dagger}$
					&.&{145} &1.3&{$\times10^{-6}$} &.&{145}
 &12.&{2}   &.&{176}  &3.5&{$10^7$} &&{$10^{6}$}  &2.5&{$\times 10^{5}$}  \\
$[$MHz$]$	&&{}    &&{}    &&{}    &&{}    &&{}    &&{}  &&{} &&{}  &&{} \\
%\hline
 &&{}    &&{}    &&{}    &&{}    &&{}    &&{} &&{}  &&{}  &&{} \\
{{$\frac{\Delta\nu_{1S-2S}}{\delta \nu_{1S-2S}}$}}
  &9.5&{$\times10^{8}$}    &1.7&{$\times10^{10}$}    &1.9&{$\times10^{15}$}
  &1.7&{$\times10^{10}$}	&2.0&{$\times10^8$}    &2.6&{$\times10^{12}$} &1.7&{$10^4$} 
 &&{$10^2$}
 &&{$10^2$} \\
 &&{}    &&{}    &&{}    &&{}    &&{}    &&{}  &&{} &&{}  &&{} \\
\hline
 &&{}    &&{}    &&{}    &&{}    &&{}    &&{} &&{}  &&{}  &&{} \\
{{$\Delta\nu _{HFS}$}}
 &20&{3.4}    &4.&{463}    &1.&{420}    &4.&{466}
 &-&{-}	&4.42&{$\times10^7$}   &-&{-}  &-&{-}  &2.1&{$\times 10^{-6}$} \\
$[$GHz$]$ &&{}    &&{}    &&{}    &&{}    &&{}    &&{}  &&{} &&{}  &&{} \\
%\hline
 &&{}    &&{}    &&{}    &&{}    &&{}    &&{}  &&{} &&{}  &&{} \\
{{$\delta\nu _{HFS}$}}
&12&{00}    &.&{145}    &4.5&{$\times10^{-22}$}    &.&{145}
&-&{-}    &.&{145} &-&{-}  &-&{-} &2.5&{$\times 10^5$} \\
$[$MHz$]$  &&{}    &&{}    &&{}    &&{}    &&{}   &&{}   &&{} &&{}  &&{} \\
%\hline
 &&{}    &&{}    &&{}    &&{}    &&{}    &&{}  &&{} &&{} &&{} \\
{{$\frac{\Delta\nu_{HFS}}{\delta \nu_{HFS}}$}}
  &1.7&{$\times10^2$}    &3.1&{$\times10^4$}    &3.2&{$\times10^{24}$}
  &3.1&{$\times10^4$}    &-&{-}    &3.1&{$\times10^8$} &-&{-}  &-&{-}  &8&{$\times 10^{-12}$} \\
 &&{}    &&{}    &&{}    &&{}    &&{}    &&{}  &&{} &&{}  &&{} \\
\hline
 \end{tabular}
%\hspace{2mm}
%
%$^\dagger${\footnotesize Only the 1S-2S splitting in the triplet
%system of positronium is considered in this table.}
}
\end{table}
 
\clearpage

\noindent
standard theory calculations are confronted with precision experiments.
In the past 50 years the chain of three natural isotopes has been significantly expanded with
artificially created ''exotic'' atoms, which often allow to study aspects of fundamental
interactions more precisely than the natural atoms.

The H-like atoms fall
in three groups: (i) purely leptonic systems such as postitronium ($e^+e^-$)  and muonium ($\mu^+
e^-$), (ii) systems containing  leptons and hadrons, such as the natural H ($pe^-$), 
deuterium ($de^-$) and tritium ($te^-$),
 but also antihydrogen ($\overline{p}e^+$), pionium ($\pi^+e^-$), muonic hydrogen ($p\mu^-$),
the muonic helium atom ((He$^++ \mu^-$)$e^-$) and antiprotonic helium atom
(He$^{++} \overline{p}e^-$) and (iii)  purely hadronic systems such as 
pionic hydrogen ($p \pi^-$) , 
the antiprotonic helium ion (He$^{++} \overline{p}$)$^+$
and protonium  ($p \overline{p}$) .
Although in principle three-body systems,  the atoms of muonic helium and antiprotonic
helium may be regarded as hydrogen-like, as one may consider the  (He$^{++} \mu^-$) 
and (He$^{++} \overline{p}$)  bound systems as pseudo-nuclei orbited by the lighter ($e^-$).
It should be mentioned that  the antiprotonic helium atom with the $\overline{p}$ in an excited state
shows both properties of an atom and of a molecule and therefore has been
named an ''atomcule'' \cite{Yamazaki_2002}.

\subsection{Positronium}
The positronium atom  (PS) as a particle-antiparticle bound system is its
own antiatom. This causes that in the theoretical description virtual
annihilations play an important role and cause significant level shifts.
PS has been formed in gases, powders and at 
certain metal surfaces. The development of sources providing
the atoms at thermal velocities in vacuum or in the
vacuum of the intergranular regions in powders has taken several decades
after the  discovery of the atom \cite{Deutsch_1951}.
This provided a base for later precision experiments.
Electromagnetic transitions have been measured in PS with 
microwave spectroscopy such as 
the ground state hyperfine splitting and  the fine structure transitions in
the first excited state of the triplet system. All are in good agreement with 
standard theory. The 1$^3$S$_{1}$ and 1$^3$S$_{1}$ 
energy difference has been determined with Doppler-free laser spectroscopy
to 2.5 ppb. This has been interpreted as the best test of 
the equality of electron and positron masses \cite{Chu_1992}.  
 
\begin{figure}[h]
  \includegraphics[height=.35\textheight,clip]{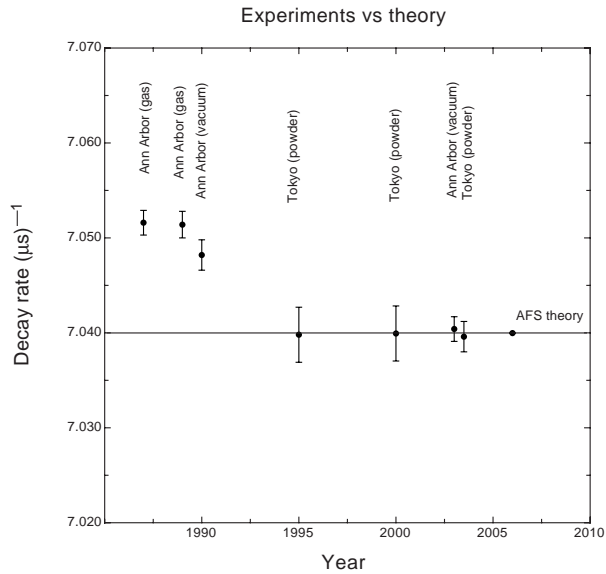}
  \caption{\label{PS} The history of orthopositronium lifetime measurements (from \cite{Rubbia_2004}).}
\end{figure}

The annihilation of triplet positronium into three gammas has been 
reported to deviate from SM calculations over the last decade.
Recent measurements\cite{Vallery_2003}  are in good agreement with theory
\cite{Rubbia_2004} (see Figure \ref{PS}).
Apparently systematic errors had been underestimated in the past.
Sensitive searches  for rare positronium decays, e.g. decays into C-parity violating numbers of 
photons or into new particles such as axions, have not given any result which would be inconsistent with
the SM. More sensitive searches are underway and motivated by a number of speculative models
\cite{Rubbia_2004} .

\subsection{Muonium}

Muonium (M)  consists of an antiparticle ($\mu^+$)  and a particle ($e^-$) \cite{Jungmann_2004}
and is therefore
to be considered  in part as antimatter. This fact causes some principle differences in the interactions
in the bound state compared to H. Well understood are the differences caused (i) by the
the different muon and proton masses , which lead to different reduced mass effects, 
(ii) by the different anomalous magnetic moments reflecting the significantly different muon and 
proton g-factors due to the proton's internal structure, and (iii) by different signs in the weak interaction 
contributions to level energies, because the proton is a particle and the muon is an antiparticle.

\begin{figure}[t]
 \includegraphics[height=.305\textheight]{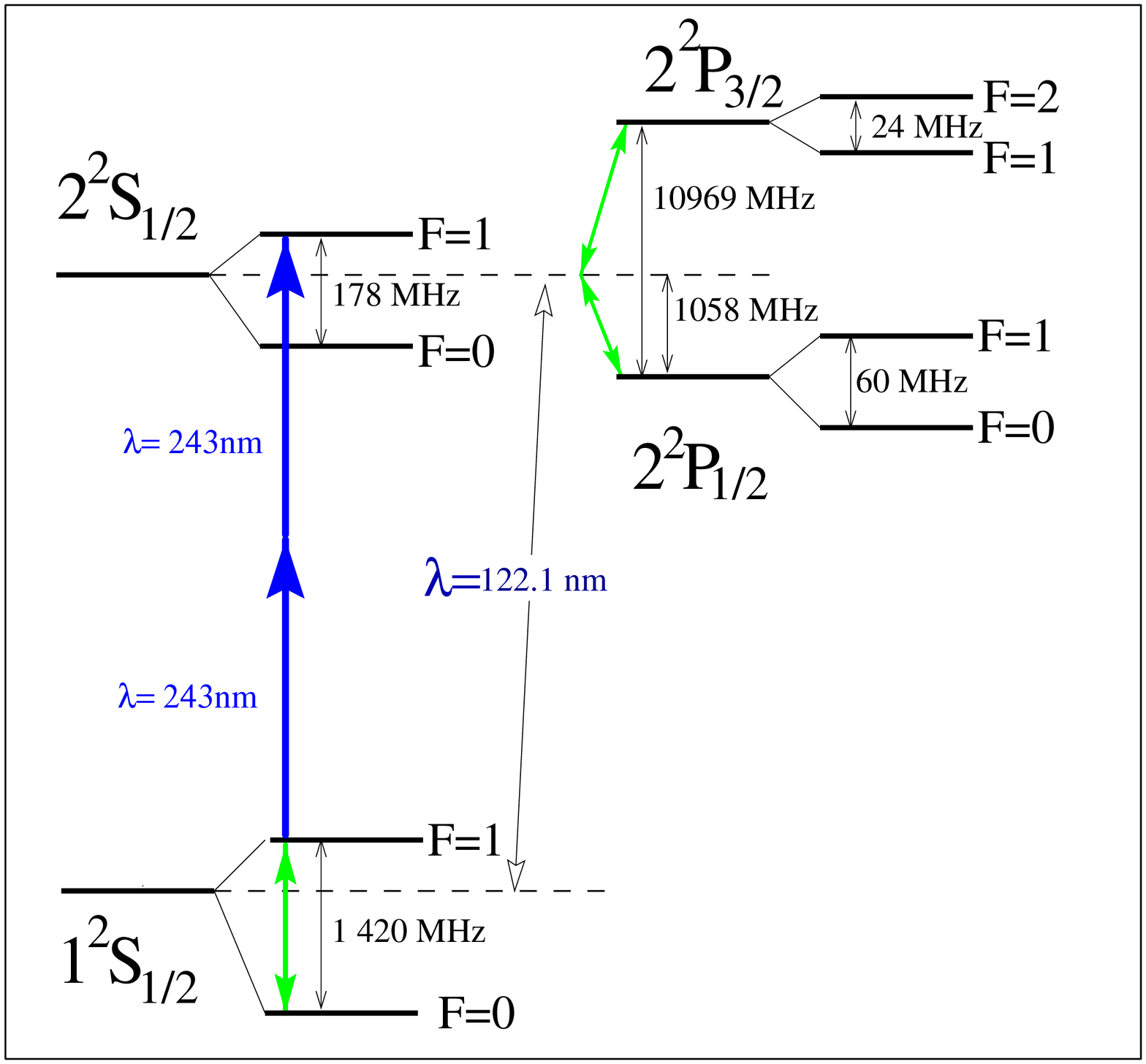}
\fbox{
 \includegraphics[height=.3\textheight]{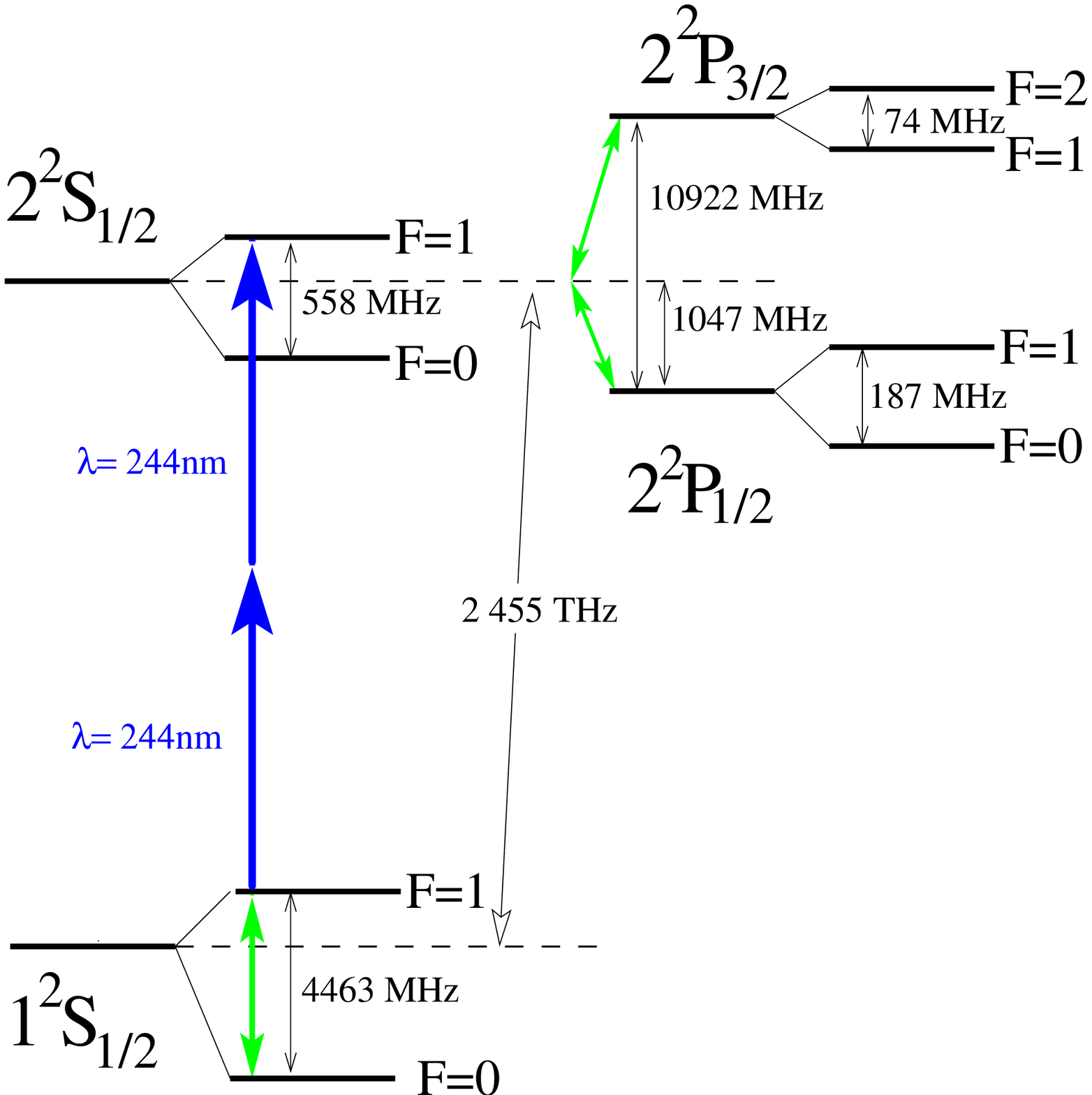} 
           }
 \caption{\label{M_term} 
           The energy levels of H and $\overline{{\rm H}}$ (right) and M (left)
           are identical and standard theory predicts exactly the same
           energy differences. Muonium energy levels differ mainly because of 
           the differences in the proton and muon masses and magnetic moments. At a smaller scale
           a difference arises from proton internal hadronic structure. New and yet unknown interactions
           such as CPT or lepton number violation forces could cause additional small energy shifts,
           which could be revealed in precision experiments.{\it (not to scale)}}
\end{figure}

%\subsubsection{Atom Formation}
In  the early years muonium research concentrated on
measurements that were possible with
atoms created by stopping muons in a gas
and studying them in this environment  \cite{Hughes_1960}.
All high precision experiments in M up to date atom 
have involved the 1s ground state (see Fig.\ref{M_term}),
in which the atoms can be produced in sufficient quantities. 
M at thermal velocities in vacuum can be obtained
by stopping $\mu^+$ close to the surface of a SiO$_2$ powder target, where 
the atoms are formed through $e^-$ capture and some of them diffuse through 
the target surface
into the surrounding vacuum, or from hot metal surfaces.
This process has an efficiency of a few percent. 
Only moderate
numbers of atoms in the metastable 2s state can be produced
with a beam foil technique. 
Because furthermore these atoms have keV energies due
to a velocity resonance in their formation.
Electromagnetic transitions in excited states, particularly 
the n=2  fine structure and Lamb shift 
splittings could be induced by microwave spectroscopy.  
However, the experimental accuracy is 1.5~\% ,
which represents not a severe test of theory. 
The detailed work over 3 decades since the discovery
of M in order to understand the atom formation resulted in
improvements of  the efficiency and quality of M  sources.  It was indispensable
for the success of novel precision experiments \cite{Jungmann_2004}.

%\subsubsection{Ground State Hyperfine Structure}

For the M ground state hyperfine 
 structure splitting the constant improvements in experimental techniques
 resulted in a factor of 10 gain in the accuracy every six years
 for a period of 20 years after the atom had been discovered. Today the experiments
 are limited by the available muon fluxes. This is the main reason why the
accuracy improvement became slower. The latest  measurements of the M 
hyperfine structure  which yield
important fundamental constants such as the muon magnetic moment
or $\alpha$   were performed at LAMPF, the brightest quasi-pulsed muon source, 
which delivered $10^8$ $\mu^+$/s. The experiment  reached 12 ppb accuracy 
\cite{Liu_1999} and was only limited by statistics.

\begin{figure}[b]
%\unitlength 1.0cm
%\begin{minipage}{5.0cm}
%\begin{picture}(6.0,6.0)
%\mbox{
  \includegraphics[height=.34\textheight,clip]{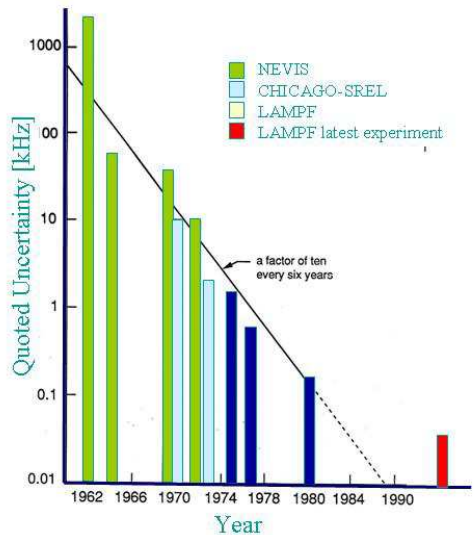}
%            }
%\end{picture}
%\end{minipage}
%\hfill
%\begin{minipage}{3.0 cm}
%\mbox{
 \caption{The accuracy of M $n=1$ state hyperfine structure measurements
                   over four decades. }
%              }
%\end{minipage} 
\end{figure}

%\subsubsection{Muonium-Antimuonium-Conversion}
The process of muonium to antimuonium-conversion 
(${\rm M}$-$\overline{{\rm M}}$)violates additive
lepton family number conservation. It would be an analogy in the lepton sector 
to  the well known $K^0$-$\overline{K^0}$ and 
$B^0$-$\overline{B^0}$ oscillations in the quark sector.
As the oscillations are hindered by gas collisions, the
availability of M in vacuum was essential for significant 
progress. When thermal M atoms in vacuum from a SiO$_2$ powder target, 
could be  observed for  their decays in an apparatus covering a large solid angle,
three and a half orders of magnitude improvement were possible for the
limit on the conversion probability which led to severe restrictions for several
speculative models \cite{Willmann_1999}. 
Again these results are limited by statistics, i.e. the available 
particle fluxes.

%\subsubsection{Laser Spectroscopy of Muonium}

With the availability of thermal M atoms in vacuum from either SiO$_2$ powder
or hot metal surfaces, laser spectroscopy became possible. It took one decade from the 
pioneering Doppler-free 1S-2S two-photon transitions at KEK \cite{Chu_88} to
the precision experiment at RAL \cite{Meyer_2000} where 4 ppb accuracy was achieved
which led to the best test of the equality of muon end electron charges at 2 ppb.
Some 15 years after the first laser experiment in M a novel technique succeeded
to obtain ultraslow polarized muons through resonant photo ionization of M
\cite{Matsuda_2003}.
Very recently a first muon spin rotation signal with muons from such a source was reported. 
This development is very important for surface and thin film science.

%\subsubsection{Test of CPT-Invariance}

\begin{figure}[h] 
\includegraphics[height=0.22\textheight,clip]{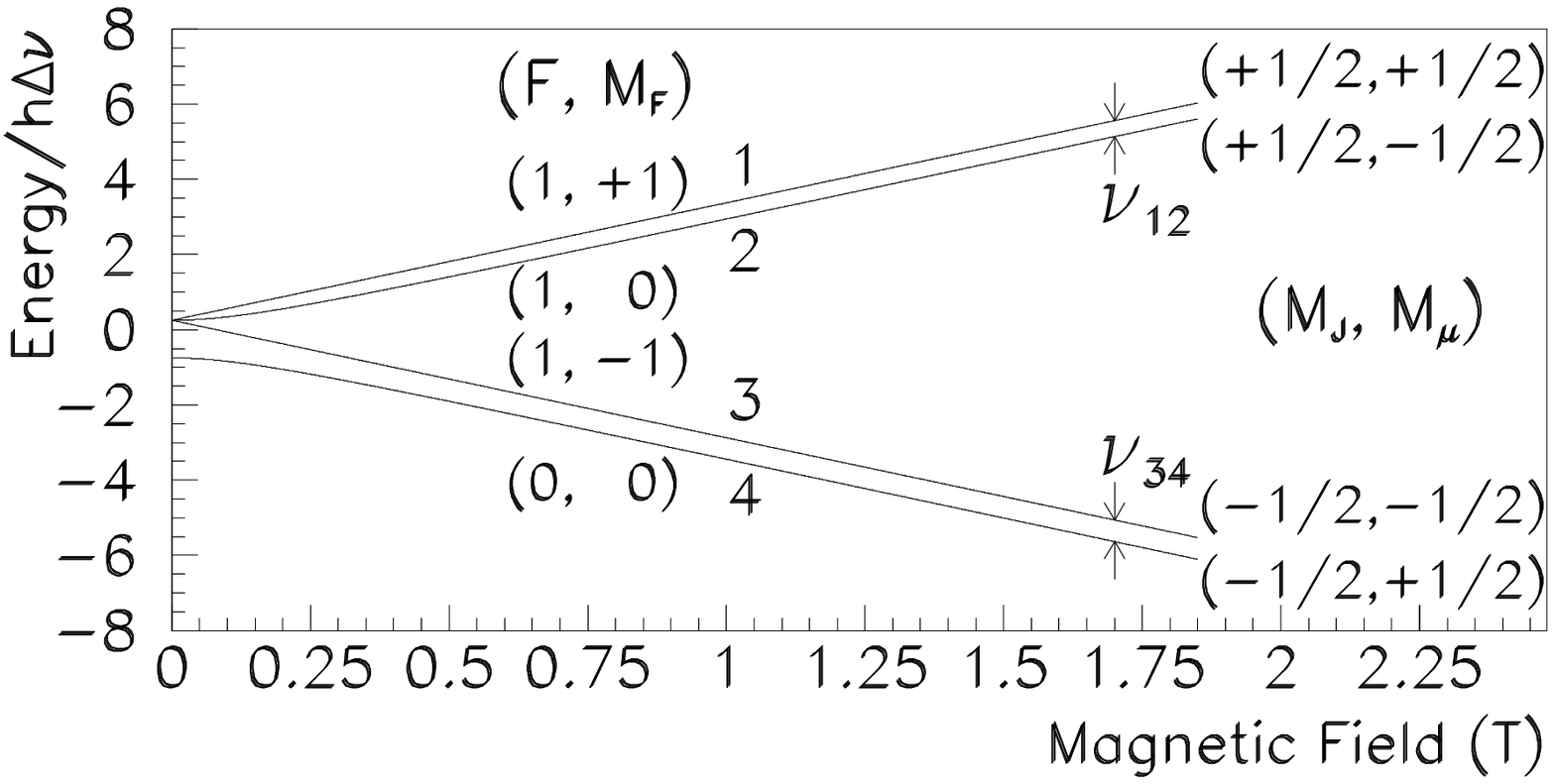}  
\includegraphics[height=0.27\textheight,clip]{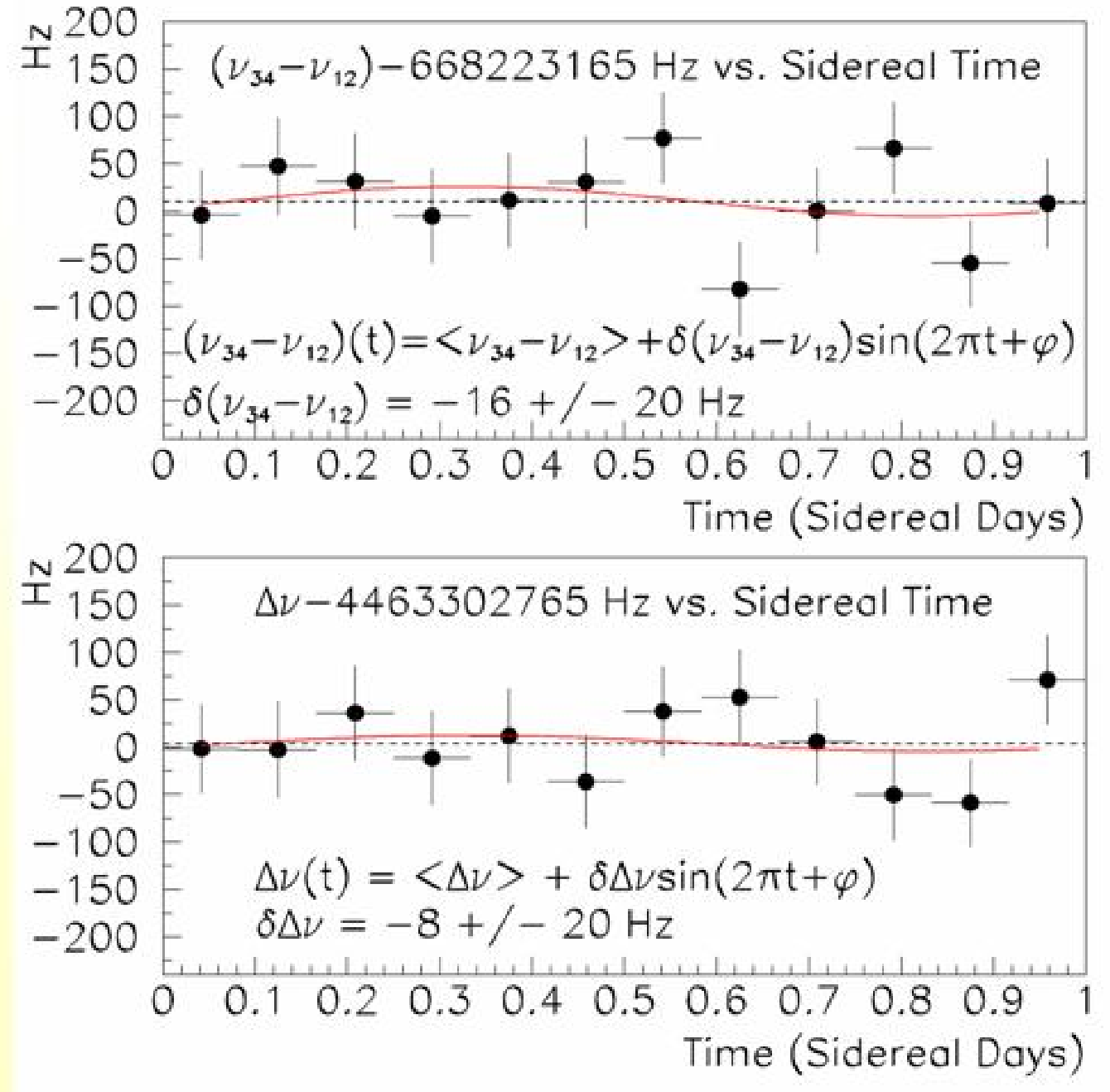}  
\caption{ \label{CPT_fig}
Left: Breit Rabi diagram of the M ground state. Right: The absence of a 
significant siderial oscillation in the ratio of transition frequencies
between Zeeman levels in a 1.7 T magnetic field
confirms CPT invariance at the best level tested for muons. }
\end{figure}

Precise CPT tests from comparing particle and antiparticle properties (see Fig. \ref{CPT_tests})
neglect in part the fact that symmetry violations ought to be connected with an interaction.
Furthermore, the CPT theorem relates to many more intriguing features in physics than
particle properties. Recently, generic extensions of the Standard Model, 
in which both Lorentz invariance and CPT invariance are not assumed, have 
attracted widespread attention in physics. In particular, new approaches try to
quantify CPT violation through additional terms in the Lagrangian of a system 
\cite{Lehnert_2005}.
For M diurnal variations of the ratio 
($\nu_{12} - \nu_{34})/(\nu_{12} + \nu_{34}$) could be a sign of Lorentz and CPT violation 
(see Fig. \ref{CPT_fig}). 
An upper  limit can be set at 
$2\cdot 10^{-23}$ GeV for the relevant parameter. 
In a specific  model by Kostelecky and co-workers a dimensionless figure of merit for CPT 
tests is sought by normalizing this parameter to the particle mass. In this 
framework $\Delta \nu_{HFS}$ provides a significantly better test of 
CPT invariance  than electron g-2 and the neutral Kaon oscillations\cite{Hughes_2001}. 
\vspace{-0.2cm}

\subsection{Hydrogen}

Atomic H has not only enormously contributed to the development of
modern physics, in particular to quantum mechanics and 
QED. Spectroscopic techniques have been pushed forward in
attempts to gain ever higher accuracy. Microwave techniques around the
H maser and laser techniques and Doppler-free spectroscopic methods
are among the examples.

%\subsubsection{Ground State Hyperfine Structure}

In atomic H the ground state hyperfine structure splitting at 1.42 GHz is
used in H masers. It's high quality factor (ratio of transition frequency to 
natural linewidth) of $3.2\times 10^{24}$ makes it interesting for a secondary 
frequency standard. The reproducibility of experimental setups, e.g. H masers,
provides the major limitation here to a long term stability of order $10^{-14}$.
The transition frequency itself has been measured (compared to the Cs frequency 
standard)  to 0.006 ppb \cite{Essen_1971}. However, the accuracy of 
bound state QED calculations is limited  to the ppm level due to the
proton's internal structure which is not well enough known for calculations of similar accuracy.
The main uncertainty arises from the distribution of magnetism within the proton,
i.e. its magnetic radius.

%\subsubsection{1S-2S Energy Interval}

Doppler-free two-photon spectroscopy in a cryogenically cooled  atomic beam of H  
 yielded the 1S-2S energy interval  with a relative uncertainty of $1.8 \times 10^{-14}$ \cite{Niering_2004}.
The coldest H beam one can imagine at this time results from 
two-photon laser extraction from a H Bose condensate \cite{Willmann_2004, Willmann_2004a}.
Typical parameters for the H 1S-2S transition  are: 
a saturation intensity of $I_s=0.9 W/cm^2$;
an excitation rate of $R_e = 4\pi\times 84 \times (I/W/s*cm^2)^2/\Delta \nu_{exp}/Hz$,
where I is the laser intensity  and $\Delta \nu_{exp}$ is the actual observed  linewidth of the 
transition; 
a 2S state photo-ionization rate $R_p=9\times I/W/s*cm^2$;
a Zeeman shift of $\delta \nu_Z = 93 \times B ~kHz/T$, with a magnetic field B;
an ac-Stark shift of  $\delta \nu_{ac}= 1.7 \times I ~Hz/W*cm^2$.
At 1 mK temperature the average velocity of the atoms is $4 m/s$ causing in a typical 600 $\mu m$ 
diameter beam  a time-of-flight broadening of $\Delta \nu_{exp}= \Delta \nu_{TOF} = 3 kHz$.
Assuming that a 1S-2S transition is observed with an MCP  after field quenching of the 2S state
a $10^{-6}$ overall detection efficiency (intrinsic efficiency $\times$  solid angle) must be taken 
into account. 
For $10^{11}$ atoms  of 30 $\mu$K in a trap a relative uncertainty for the  line center
of $\delta \nu / \nu_{1s-2s} = 10^{-13}$ has been achieved in 1 s integration time 
\cite{Willmann_2004a}. 
(It should be noted that at saturation intensity the ionization rate is of order 1.7 s$^{-1}$ and that
the atoms can be used essentially only once, because after the transition the $m_F$ states get
 equally populated.)
In the absence of  systematic effects  the uncertainty of the line center scales as
\begin{equation}
\label{H_accuracy}
\delta \nu = \Delta \nu_{exp}/(Signal/Noise) = \Delta \nu_{exp}/\sqrt{N}~~,
\end{equation}
where $N$ represents  the number of observed transitions which is in  approximation 
proportional to the number of atoms in the laser field.
The reachable precision depends strongly 
on the number of available atoms.
\vspace*{-0.1cm}
\begin{figure}[t]
  \includegraphics[height=.265\textheight,clip]{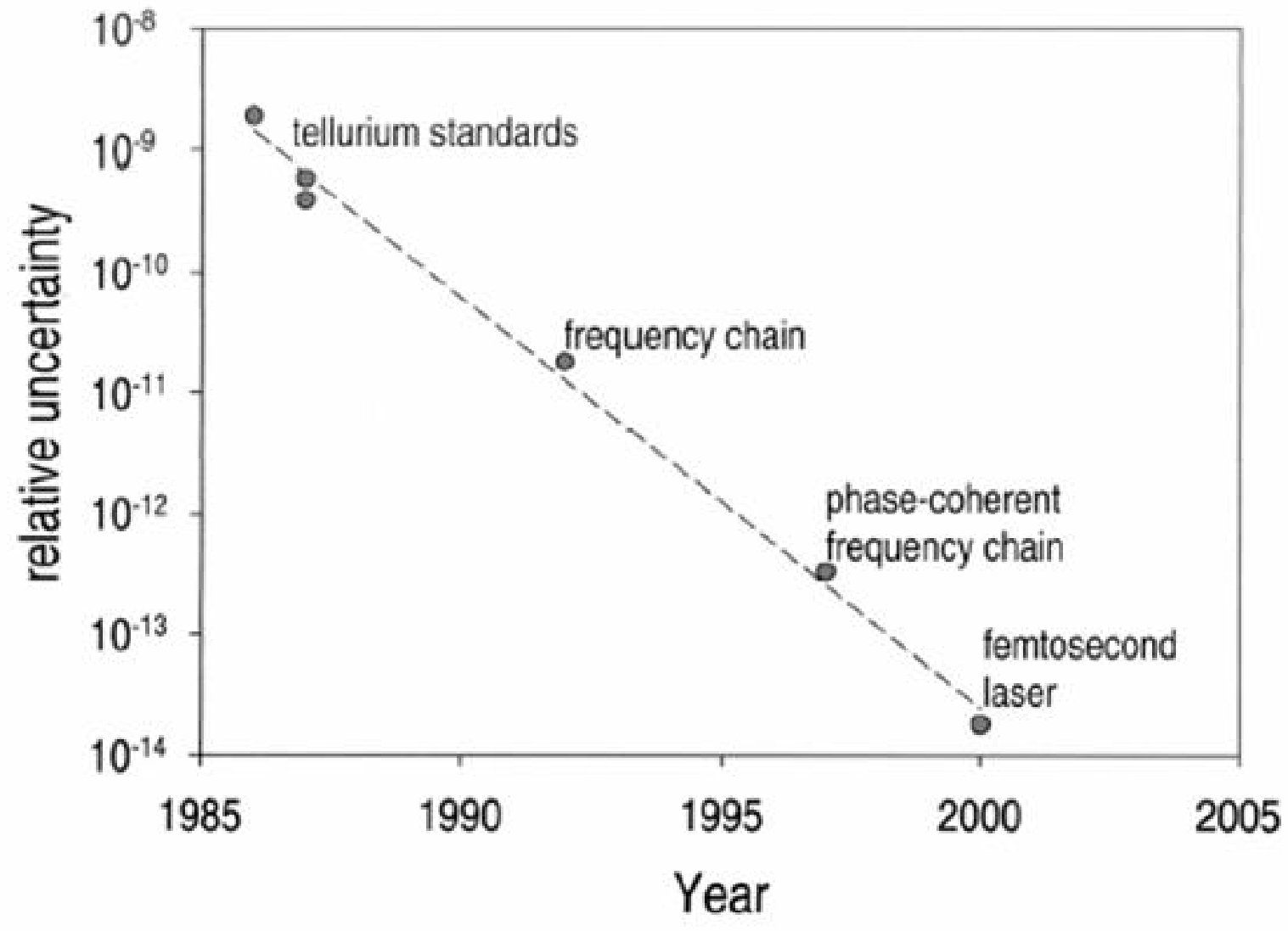}
  \hspace*{-.5cm} 
  \includegraphics[height=.265\textheight,clip]{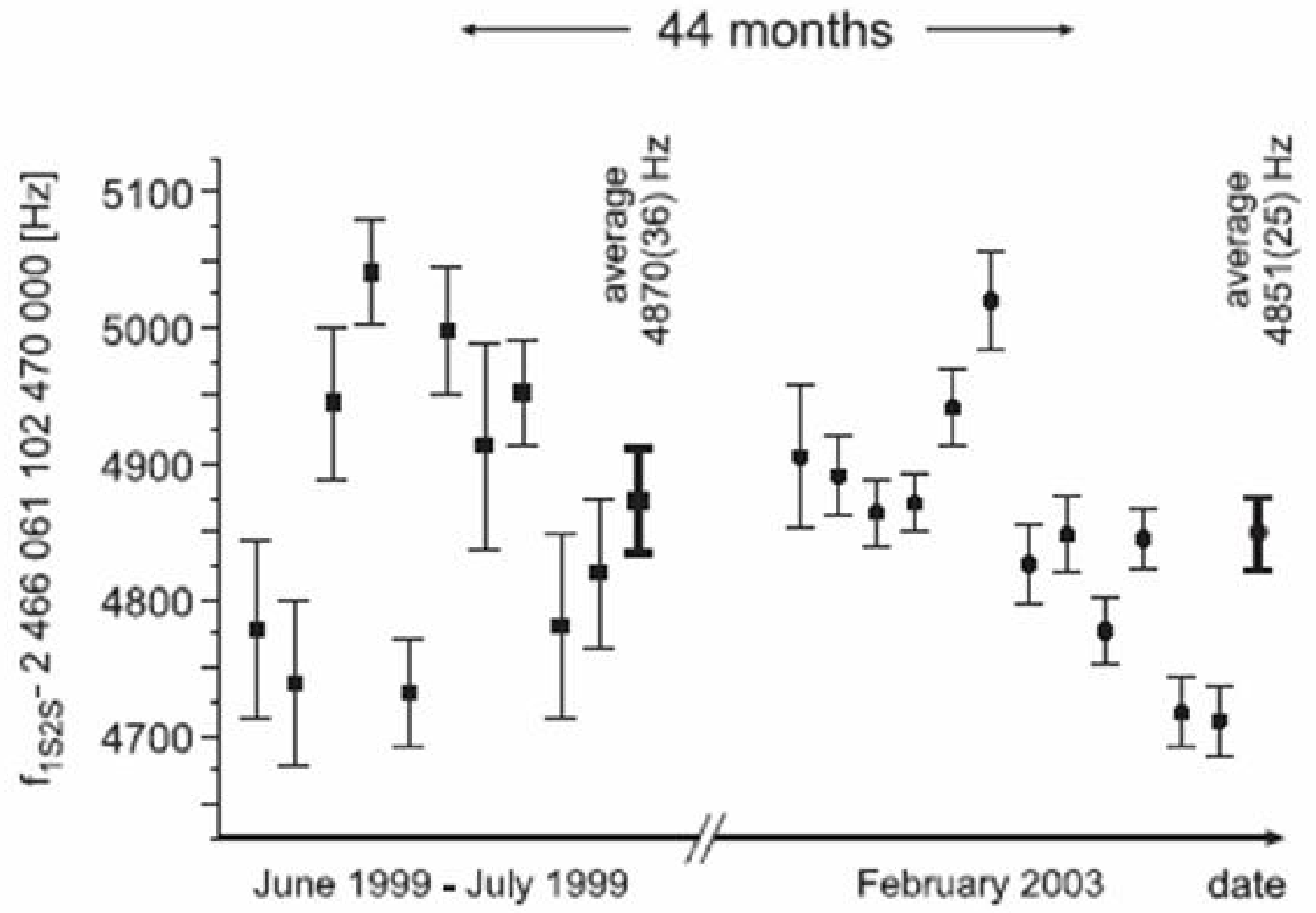}
  \caption{Right: Measurements of the 1S-2S transition frequency in atomic H
           have been reported with continuously improving relative uncertainty 
          down to  the $10^{-14}$ level 
          due to ever  improving technology.
                   Left: Two series of precision measurements of the 1s-1s energy
           difference in H over a4 years time  allow
           to extract a limit on the time variation of
           the finestructure constant \protect{$\alpha$} \cite{Fischer_2004}.
                 }
\end{figure}

%\subsubsection{Variation of Finestructure Constant in Time}
 A limit on the time variation of \protect{$\alpha$} was extracted from
two series of repeated measurements of the 1s-2s energy
difference in H over a long time and 
%\begin{equation}
$\frac {\partial \alpha}{\partial t} / \alpha = \frac{\partial}{\partial t}(\ln \alpha)
 = (-0.9\pm2.9)\times 10^{-15} y^{-1} $ 
%\end{equation}
could be established \cite{Fischer_2004}.
We note, the two series  have reduced $\chi^2$ values of 4.2 respectively 9, 
which may be viewed as a hint that
 present experiments suffer from not well understood 
 systematic errors.  Therefore, presently optical spectroscopy
appears to be limited at the \protect{$10^{-13}$} level of 
relative accuracy.

\subsection{Antihydrogen}  
\subsubsection{Antihydrogen Formation}
$\overline{{\rm H}}$ was produced first  at CERN in 1995
\cite{Baur_1996} . The atoms were fast as the production mechanism required $e^+e^-$ 
pair creation when  $\overline{p}$'s were passing near heavy nuclei. A small
fraction of the $e^+$ form a bound state with the $\overline{p}$.
The experiment was an important step forward showing that  a few $\overline{{\rm H}}$
could be produced.
Unfortunately, the speed of the atoms does not allow any
meaningful spectroscopy. Later  a similar experiment was carried out at FERMILAB 
\cite{Blanford_1998}.

The successful production of slow $\overline{{\rm H}}$ was first reported by the
ATHENA collaboration in 2002 \cite{Amoretti_2002} and shortly later also by the ATRAP collaboration
\cite{Gabrielse_2002}. Both experiments use combined Penning traps in which first
$e^+$  and $\overline{p}$ are stored separately and cooled. The atoms form when both species 
are brought into contact by proper electric potential switching in the combined traps.
The detection in ATHENA  relies on diffusion of the neutral atoms out of the interaction volume and
the registration of $\pi$'s  which appear when the  atoms annihilate on contact with matter walls of the
container. In ATRAP the hydrogen atoms are re-ionized in an electric field 
and the $\overline{p}$'s are observed using 
a capture Penning trap. 

Most of the atoms are in excited states (n > 15) which can be seen from the fact that their physical size
is above 0.1 $\mu$m\cite{Gabrielse_2004}. 
For spectroscopy the atoms need to be in 
states with low $n$, 
preferentially the ground state. The production of such states is a major goal 
of the community for the immediate future.
The kinetic energy of the produced $\overline{{\rm H}}$ atoms has been measured to be of
order 200 meV corresponding to a velocity of $6 \times 10^4$ m/s . 
This is a factor of 400 above the value where neutral atom traps can hold them.
Therefore cooling such atoms or identifying a production mechanism for colder 
$\overline{{\rm H}}$ are a  central topic.

For laser cooling of $\overline{{\rm H}}$  a continuous laser at the H Lyman-$\alpha$
frequency for
$\overline{{\rm H}}$ cooling  has recently been developed \cite{Eikema_2001}. 
One hopes to 
achieve  the photo-recoil limit of 1.3 mK.

Recently a promising new method was demonstrated to obtain antihydrogen. It uses resonant 
charge exchange with excited positronium to obtain $\overline{{\rm H}}$ atoms with 
essentially the same velocities as the $\overline{p}$'s in the trap which can
be made rather low by cooling. \cite{Storry_2004,Gabrielse_2004}. 

\begin{figure}[b]
  \includegraphics[height=.25\textheight,clip]{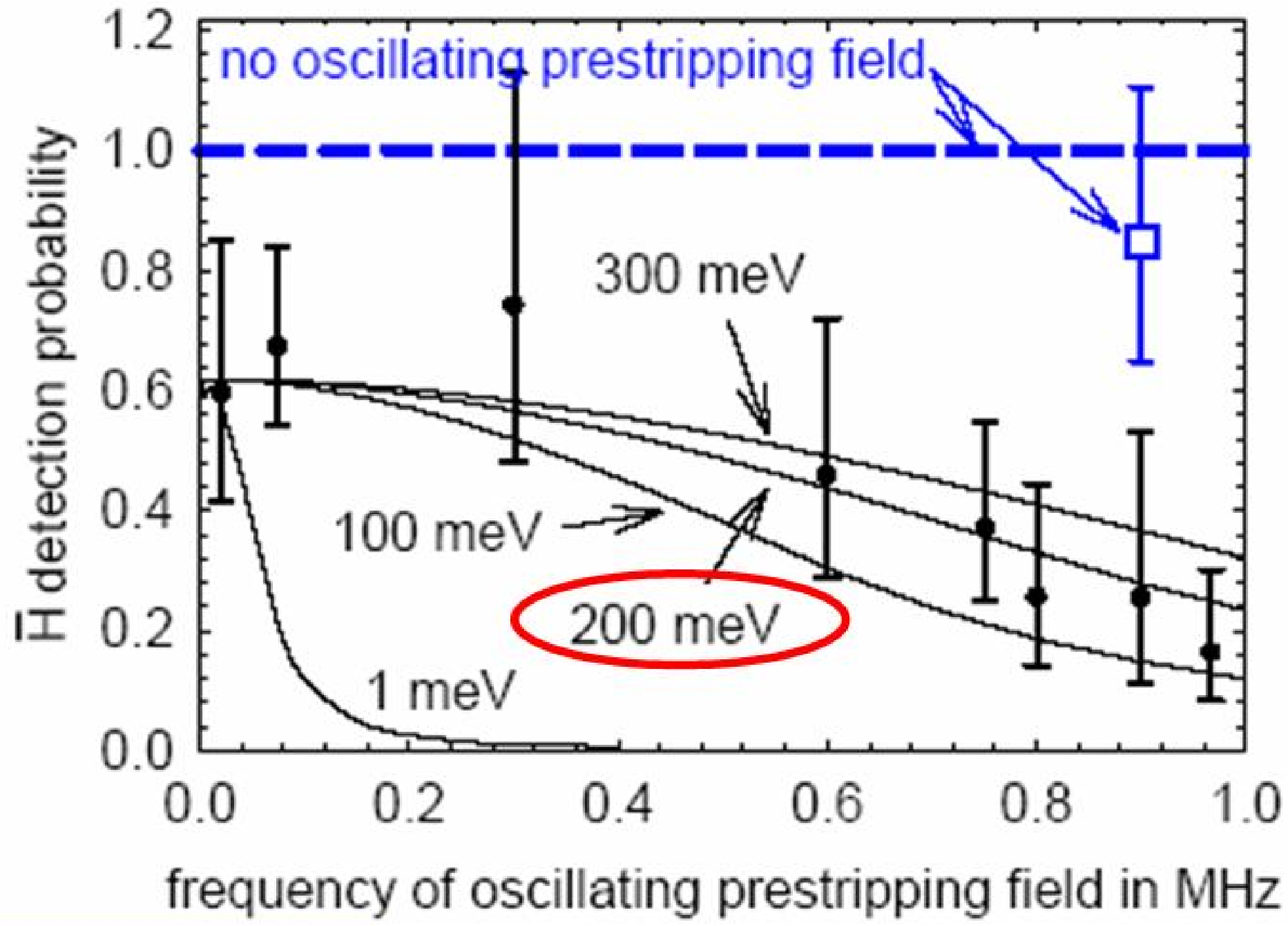}
  \includegraphics[height=.25\textheight]{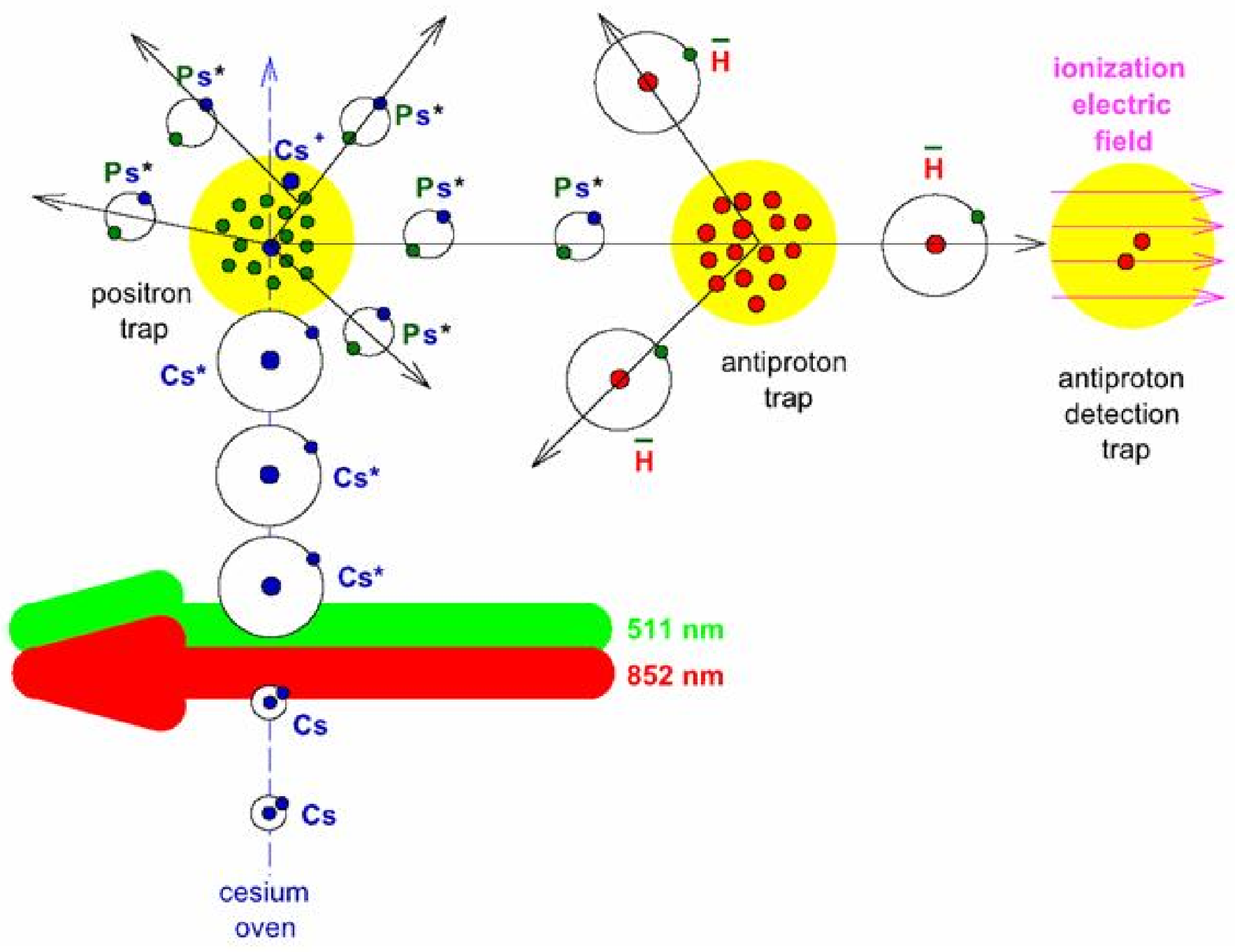}
  \caption{Right: The $\overline{{\rm H}}$ atoms produced up to date have 
                  typical kinetic energies of order 200meV. Kinetic energies 
                  below 0.5 meV are needed to trap the atoms. Left:
                  A new method to produce $\overline{{\rm H}}$ uses charge 
                  exchange with PS atoms. \cite{Gabrielse_2004}
                     }
\end{figure}

\subsubsection{CPT Tests with Antihydrogen}
A main motivation to perform precision spectroscopy on $\overline{{\rm H}}$
is to test  CPT invariance.  There are two electromagnetic transitions which offer a
high quality factor (see Table \ref{tab_H}) and therefore promise high experimental precision
when H and $\overline{{\rm H}}$ are compared:
the 1S-2S two-photon transition at frequency $\Delta \nu_{1s-2s}$ 
and the ground state hyperfine splitting $\Delta \nu_{HFS}$, which both have within the SM
in addition to the leading order contributions from QED, nuclear structure, weak and strong interactions,
\begin{equation}
\label{CPT1s2s}
\Delta \nu_{1s-2s} = \frac{3}{4} \times R_{\infty} + \varepsilon_{QED}
                                      + \varepsilon_{nucl} + \varepsilon_{weak} + \varepsilon_{strong}
                                      + \varepsilon_{CPT} 
\end{equation}
\begin{equation}
\label{CPTHFS}
\Delta \nu_{HFS} = const\times \alpha^2 \times R_{\infty} + \varepsilon_{QED}^*
                                      + \varepsilon_{nucl}^* + \varepsilon_{weak}^* + \varepsilon_{strong}^*
                                      + \varepsilon_{CPT}^*.
\end{equation}
In eq.(\ref{CPT1s2s})  and eq.(\ref{CPTHFS})  it is assumed that only CPT violating contributions
exist from interactions beyond the SM. If one assumes that $\varepsilon_{CPT}$ and
$\varepsilon_{CPT}^*$ are of the same order of magnitude, there relative contributions is larger by order
$\alpha^{-2}$  for $\Delta \nu_{HFS}$. Further one can speculate that a new 
interaction may be of short range (contact interaction), which also favors 
measurements of $\Delta \nu_{HFS}$. Such an experiment has been recently proposed.
It utilizes a cold $\overline{{\rm H}}$ atom beam and has sextupole state selection magnets in a 
Rabi type  atomic beam experiment \cite{Widmann_2004_a}.
For both experiments eq. (\ref{H_accuracy}) governs the reachable precision, i.e. the
atoms should be as cold as possible and one should use as many as possible atoms.
%
%\cite{Widmann_2004}
%\begin{figure}
%  \includegraphics[height=.3\textheight,clip]{AH_HFS.eps}
%  \caption{Principle of a proposed $\overline{{\rm H}}$ ground state hyperfine structure experiment.
%                  The atoms are extracted as a beam from a cold neutral atom trap and spin selected
%                   in a sextupole magnet. After microwave transitions have taken place in a cavity
%                   the spin state is analyzed in a second sextupole \cite{Widmann_2004}. }
%\end{figure}

\subsubsection{Gravitational Force on $\overline{{\rm H}}$}
\begin{figure}[h]
  \includegraphics[height=.22\textheight,clip]{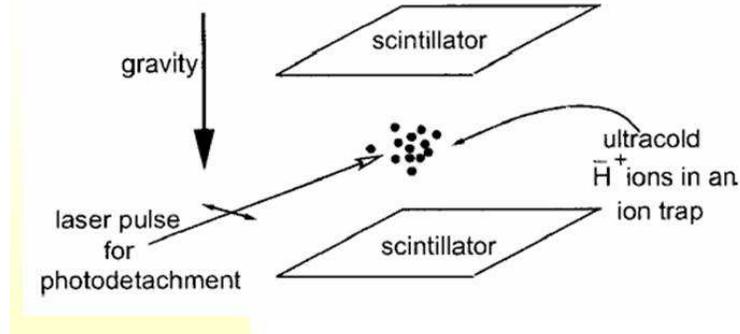}
  \caption{\label{Hbar_gravity}
           The sign of the gravitational force on atomic $\overline{{\rm H}}$
           needs to be determined by experiment, as there is 
           no experimental evidence yet that antimatter and
           matter show identical behavior concerning gravity \cite{Walz_2003}. }
\end{figure}

One of the completely open questions in physics concerns the sign of gravitational interaction
for antimatter. It can only be answered by experiment. A proposal \cite{Walz_2003} exists 
(see Fig. \ref{Hbar_gravity})  in which the deflection of a horizontal cold beam is
measured in the earth's gravitational field. The experiment plans on a number of
modern state of the art atomic physics techniques like sympathetic
cooling of $\overline{{\rm H}}^+$ ions  by ,e.g. Be$^+$ ions in an ion trap
to achieve the neccessary  low temperatures of some 20~$\mu$K . 
After  pulsed laser photo-dissociation of the ion into $\overline{{\rm H}}$  and a $e^+$ the
neutral atoms can then leave the trap. The atom's ballistic path can be measured.

\section{Antiprotonic Helium}

\begin{figure}[bt]
  \includegraphics[height=.43\textheight,clip]{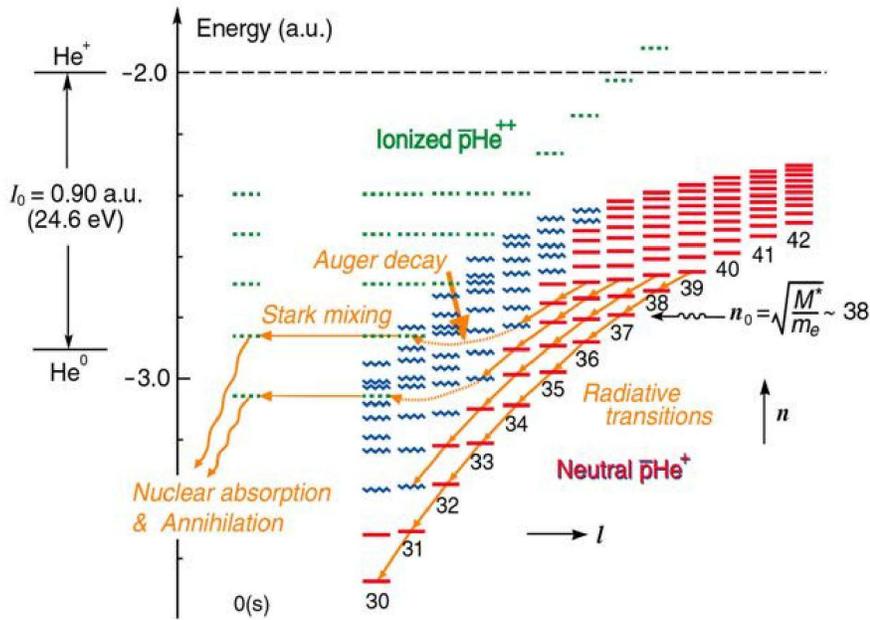}
  \caption{\label{AHe_terms}
                    In the formation process of antiprotonic helium  the capture of the $\overline{p}$
                   into a state with quantum numbers $n \approx 38$ and maximal $l$ is very likely.
                   Such states are rather stable against   $\overline{p}$ annihilation \cite{Yamazaki_2002}. 
                   }
\end{figure}

The potential of antiprotonic helium for precision measurements in the field of 
fundamental interaction research was realized shortly after it had been discovered that
$\overline{p}$'s stopped in liquid or gaseous helium exhibit long lifetimes and do not
rapidly annihilate with nucleons in the helium nucleus \cite{Yamazaki_2002}. 
This can be explained, if one assumes that the $\overline{p}$'s  are captured in metastable states
of high principal quantumnumber $n$  and high angular momentum $l$ , with $l\approx n$  (see  
Fig. \ref{AHe_terms} \cite{Nakamura_1994}. The capture happens at typically
at $n \approx \sqrt{M^*/m_e} \approx 38$ , where $M^*$ is the reduced mass of
the ($\overline{p}$He) bound system.

\begin{figure}[b]
  \includegraphics[height=.28\textheight]{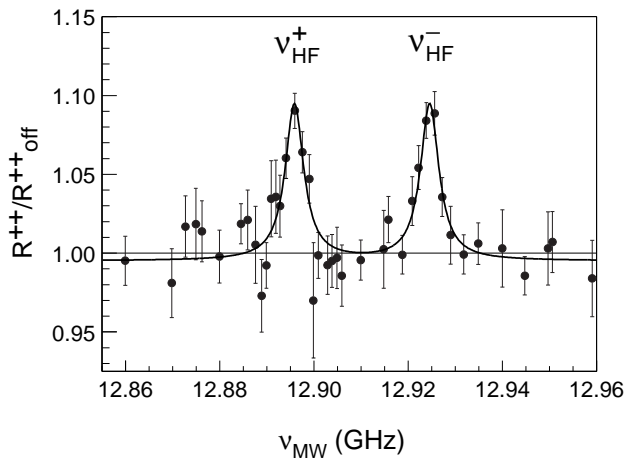}
  \caption{\label{AHe_HFS} Example of a line splitting due to hyperfine structure
                         in antiprotonic helium
                   \cite{Widmann_2002}.}
\end{figure}

With laser radiation  the $\overline{p}$'s in these atoms can be transferred into states where Auger 
de-excitation can take place. In the resulting  H-like system Stark mixing with 
s-states results in nuclear  $\overline{p}$ absorption and annihilation which is 
signaled by emitted pions. This way a number of transitions could be induced
and measured with continuously increasing accuracy over the past decade.
A precision of $6 \times 10^{-8}$ has been reached for the transition frequencies 
\cite{Hori_2001}, which has been stimulating for improving three-body QED 
calculations.  It should be noted that with the high principal quantum numbers
for the $\overline{p}$ the system shows also molecular type character
\cite{Yamazaki_2000}.

\begin{figure}[t]
  \includegraphics[height=.3\textheight,clip]{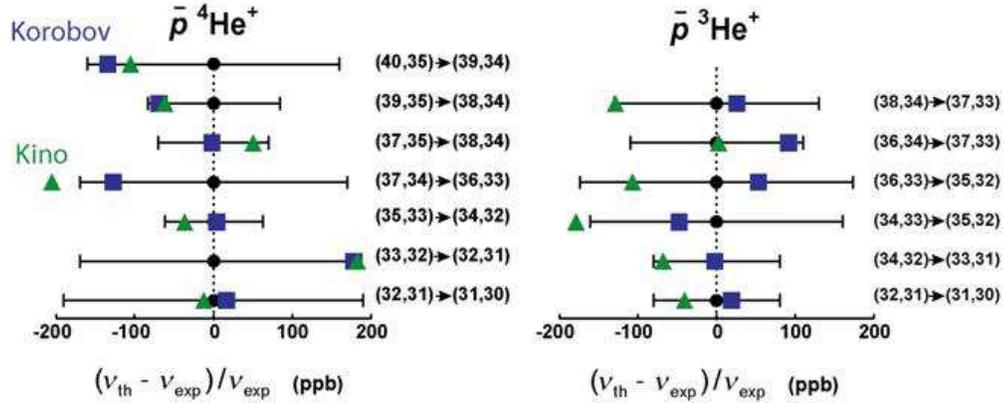}
  \caption{Accurate QED calculations have been performed for 
antiprotonic $\overline{P}^4$He$^+$ and  $\overline{P}^3$He$^+$. The differences between the
results from independent calculations are below the accuracy achieved in experiments. 
These differences may be regarded as a an indication of the
size of systematic uncertainties in the theoretical approaches chosen \cite{Hori_2003}.}
\end{figure}

Among the spectroscopic sucesses the laser-microwave double resonance
measurements of hyperfine splittings of $\overline{p}$ transitions could be 
measured (see Fig. \ref{AHe_HFS}) \cite{Widmann_2002}. There is greement with QED theory
\cite{Korobov_2001} at the $6 \times 10^{-5}$ level which can be interpreted as
a measurement of the antiprotonic bound state g-factor to this accuracy.
In principle, hyperfinestructure measurements in antipprotonic helium
offer the possibility to measure the magnetic moment of the $\overline{p}$.

%\subsubsection{Antiprotonic Helium Ions}

\subsubsection{CPT tests with Antiprotonic Helium Ions}
The very good agreement of the QED calculations with the
measurements of several transitions can be exploited to extract
a limit on the equality of the charge$^2$ to mass ratio for proton and $\overline{p}$.
Combined with the results of cyclotron frequency measurements \cite{Gabrielse_1999}
in  Penning traps one can conclude  that masses and charges of  proton and $\overline{p}$ 
are equal within $6\times 10^{-8}$ in full agreement with
 expectations based on the CPT theorem \cite{Hori_2003}.  The collaboration 
estimates that a test down to the 10 ppb level should be possible.

\begin{figure}[t]
  \includegraphics[height=.43\textheight]{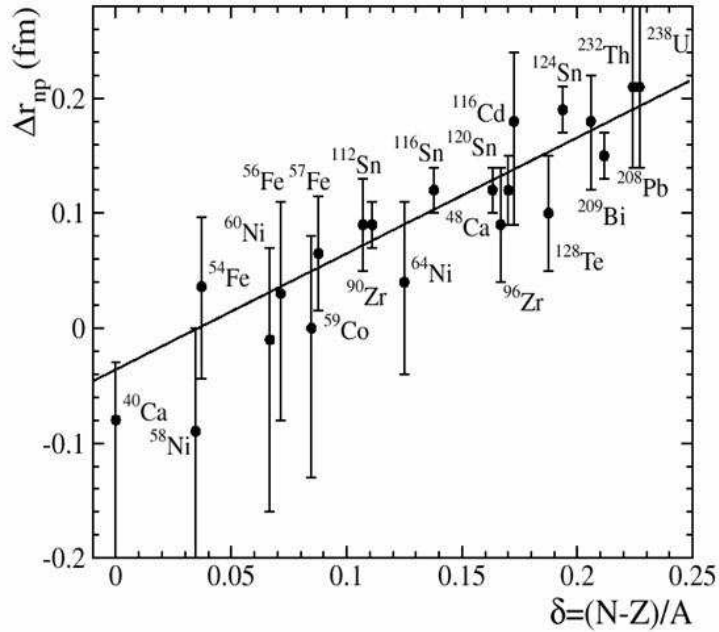}
  \caption{\label{n_radii}
                   Difference between the rms radii of the neutron and proton distributions. They were 
                   extracted from antiprotonic atom x-ray experiments \cite{Trzcinska_2004}.}
\end{figure}

\section{Antiprotonic (Radioactive) Atoms}
It has been shown at the LEAR facility at CERN
that antiprotonic x-rays from atoms in which a $\overline{p}$ has been 
captured can be utilized to obtain information on the neutron mean square 
radii of nuclei (see Fig. \ref{n_radii}) \cite{Trzcinska_2004}. The accuracy
of the experiments is limited at present by nuclear theory. Neutron distributions
are expected to be the limiting factor in the theory of the upcoming round of precision
experiments on atomic parity violation. In particular some radioactive nuclei of
Francium and Radium  isotopes are of interest. At a combined radioactive beam and
antiproton facility one can expect experiments to determine the neutron radii with
sufficient accuracy for,e.g., the needs of theory to describe  improved atomic parity violation
experiments including such in heavy radioactive atoms.

\section{Status of Slow Antiproton Physics (spring 2005) - Conclusions }
Low energy antiproton research has made already a number of important contributions
to test fundamental symmetries and to verify precise calculations. With cyclotron frequency 
measurements of a single trapped
$\overline{p}$ and  with precision spectroscopy of antiprotonic helium ions
stringent  CPT tests could be performed on $\overline{p}$ parameters. With precise measurements
of $\overline{p}$ in antiprotonic helium atomcules the bound state QED
three-body systems could be challenged, which has led to significant advances already.
 With antiprotonic heavy atoms
new input could be provided to obtain neutron radii of nuclei. The differences
in proton/ antiproton interactions with matter could expand on similar work
with other particle/ antiparticle systems.

 $\overline{{\rm H}}$ atoms have been produced
by two independent collaborations. 
Precision spectroscopy of these atoms will
depend on the availability of atoms in the ground state, 
the successful  cooling of the systems to below the 100 ~$\mu$eV range and their confinement in neutral particle traps. 
Work in this direction is in progress. A comparison with other exotic atom experimental programs
shows that one must allow for sufficient time to develop the necessary understanding
of production mechanisms and one must allow time 
for improving the techniques. The experiments will benefit in their speed of progress and in their 
ultimate precision
from future slow $\overline{p}$ sources of significantly improved particle fluxes and brightness 
as compared to
today's only operational facility. 

The ongoing and planned experiments bear a robust discovery potential for new physics,
in particular when searching for CPT violation. We can look forward to future precision 
$\overline{p}$ experiments continuing  to deepen insights in fundamental interactions 
and symmetries, providing  important data and parameters with standard theory  and 
providing  improved searches for new physics.

\section{Acknowledgements}
The author wishes to express his gratitude to the 
organizers under the leadership of Y. Yamazaki for creating a stimulating 
atmosphere during the workshop and for their support. The author is further grateful to
E. Widmann for numerous discussions on the various subjects covered and to
L. Willmann for carefully reading the manuscript.
This work was in part
supported by the Dutch Stichting voor Fundamenteel Onderzoek (FOM)
in the framework of the TRI$\mu$P programme.

\end{document}